\documentclass[a4paper,11pt]{article}
\pdfoutput=1 
\usepackage{jheppub} 
\usepackage[T1]{fontenc}

\usepackage{graphicx}
\usepackage{amsmath,amssymb,color}
\usepackage{grffile}
\usepackage{ulem}
\usepackage{bm}
\usepackage{float}
\usepackage{color}
\usepackage[caption=false]{subfig}
\usepackage{xcolor}

\newcommand{\beq}{\begin{equation}}
\newcommand{\eeq}{\end{equation}}
\newcommand{\be}{\begin{equation}}
\newcommand{\ee}{\end{equation}}
\newcommand{\ba}{\begin{eqnarray}}
\newcommand{\ea}{\end{eqnarray}}

\definecolor{fashionfuchsia}{rgb}{0.96, 0.0, 0.63}

\title{
Kink-antikink scattering in a quantum vacuum
}
\author[a,1]{Mainak Mukhopadhyay,\note{Corresponding author.}}
\author[b]{Evangelos I.~Sfakianakis,}
\author[a]{Tanmay Vachaspati}
\author[b]{and George Zahariade}

\affiliation[a]{Department of Physics, Arizona State University,\\ 450 E. Tyler Mall, Tempe, AZ 85287-1504, U.S.A.}
\affiliation[b]{Institut de F\'isica d'Altes Energies (IFAE), The Barcelona Institute of
Science and Technology (BIST), Campus UAB, 08193 Bellaterra, Barcelona}

\emailAdd{mmukhop2@asu.edu}
\emailAdd{esfakianakis@ifae.es}
\emailAdd{tvachasp@asu.edu}
\emailAdd{gzahariade@ifae.es}

\abstract{
We study kink-antikink scattering in the sine-Gordon model 
in the presence of interactions with
an additional scalar field, $\psi$, that is in its quantum vacuum. 
In contrast to the classical scattering, now there is quantum radiation of
$\psi$ quanta and the kink-antikink may form bound states that resemble
breathers of the sine-Gordon model. We quantify 
the rate of radiation and map the parameters for which bound states are formed. 
Even these bound states radiate and decay, and eventually there is a transition into 
long-lived oscillons.
}

\begin{document}

\maketitle
\flushbottom

\section{Introduction}
\label{sec:intro}


Topological defects are 
expected to arise during phase transitions in the early universe. These can include domain walls \cite{Saikawa:2020duz, Hiramatsu:2012sc, Vilenkin:1982ks, Sikivie:1982qv}, cosmic strings \cite{Vilenkin:1984ib, Blanco-Pillado:2017rnf, Dufaux:2010cf, Damour:2000wa, Albrecht:1984xv,  Vachaspati:1984gt, Hindmarsh:1994re, Kibble:1976sj}, monopoles \cite{tHooft:1974kcl, Nambu:1974zg, Gibbons:1986df, Mavromatos:2020gwk} and textures \cite{Gueron:1997ck, Notzold:1990jt, Spergel:1992ff, Durrer:1990qe, Turok:1989ai, Brandenberger:2019txi}, depending on the symmetry breaking pattern.  While topological defects have not yet been observed in a cosmological context \cite{Planck:2013mgr}, they are abundant in condensed matter systems~\cite{Zurek:1985qw,PhysRevLett.119.013902,PhysRevLett.105.075701,Chuang:1991zz,Bowick:1992rz,Hendry1994,Ruutu1996,Bauerle1996,PhysRevLett.89.080603,PhysRevLett.84.4966,PhysRevLett.91.197001,Beugnon2017}, with intriguing analogies between the two areas of research \cite{Kibble:2001wj,  Vachaspati:1998vc, Zurek:1996sj}.

Topological defects are also an ideal setting for studying quantum effects in
spacetime dependent classical backgrounds. Significant effort has already 
gone into the quantization of topological defects \cite{Rajaraman:1982is, Coleman:1985rnk, Christ:1975wt, Goldstone:1974gf, Cahill:1974cs, Dashen:1974ci, Dashen:1974cj, Dashen:1974ck}, in which case
the topological defects form space-dependent backgrounds for quantum 
fields. Less effort has been invested in time-dependent problems that involve 
quantum fields in the background of dynamical topological defects  \cite{Dashen:1975hd}. Going one step further, the quantum fields will 
backreact on the dynamics of the topological defects and lead to non-trivial
interplay between the quantum and classical degrees of freedom. This is the
subject of the present paper.

The sine-Gordon equation in one spatial dimension holds a special place in the soliton literature, 
since it is integrable and allows for the construction of multi-soliton solutions by the use of the 
B\"acklund transformation. The solitons (or kinks) of the sine-Gordon equation are solutions which 
interpolate between two successive minima of the potential.
The sine-Gordon model has been analyzed in great detail~\cite{Rajaraman:1982is,Coleman:1985rnk,Vachaspati:2006zz} and
kink-antikink scattering is known to be trivial -- the kink and antikink pass
through each other albeit with a time delay -- a result that holds even when 
$\phi$ is treated in quantum field theory. 
scattering in a model where
the sine-Gordon scalar field, $\phi$, is coupled to a second scalar field, $\psi$.
The kink and antikink propagate
in the quantum vacuum of $\psi$, deforming the vacuum around them. (Alternately, the kink and antikink get dressed by the $\psi$ field.)
Upon kink-antikink scattering, $\psi$ particles (wavepackets) are radiated. Depending on the
parameters, the quantum radiation of $\psi$ particles can lead to the production
of a kink-antikink
bound state that is known as a ``breather'' in the sine-Gordon literature~\cite{Rajaraman:1982is,Coleman:1985rnk,Vachaspati:2006zz}.
In this case, the breather oscillates and continues to radiate $\psi$ particles. (This was more specifically studied in Ref.~\cite{Olle:2019skb}, albeit in a slightly different model.)
Surprisingly we find that the breather eventually settles into another oscillating
configuration that radiates very slowly, resembling an ``oscillon''. Oscillons are long-lived localized configurations arising in a variety of scalar field theories and have been shown to emerge naturally in scenarios ranging from preheating to bubble collisions (see e.g. \cite{Amin:2011hj, Copeland:1995fq,  Gleiser:2009ys, Farhi:2005rz,  Graham:2006vy,  Zhou:2013tsa, Fodor:2008es, Hiramatsu:2020obh, VanDissel:2020umg,  Zhang:2020bec, Zhang:2020ntm}).

Our analysis is enabled by the ``classical-quantum correspondence'' (CQC) in
which the dynamics of free quantum fields in spacetime dependent classical 
backgrounds has been shown to be equivalent to a completely classical problem
in higher dimensions \cite{Vachaspati:2018hcu, Olle:2019skb, Mukhopadhyay:2019hnb}. Quantum backreaction on the classical background 
is included in the analysis in the semiclassical approximation in which quantum
operators occurring in the classical equations of motion are replaced by their
dynamical expectation values. Special attention is given to the renormalization issues that arise, since a divergent 
part of the expectation values contributes to the mass parameter of the background 
field.

In Sec.~\ref{sec:setup} we introduce the field theoretic model, as well as the equations 
that will be solved numerically to determine the full dynamics of the kink-antikink collision. 
We take care to distinguish the no backreaction case, where the kink-antikink background 
is fixed, from the general case with backreaction. In Secs.~\ref{sec:results} and~\ref{sec:cases} we outline the 
results of our numerical analysis: we determine the region of parameter space where a breather-like bound state is formed after the collision and discuss the different phases of its decay. We conclude in Sec.~\ref{sec:conclusions}, where we discuss the implications of this work, its limitations as well as several future prospects.
We use natural units where $\hbar=c=1$ throughout.

\section{Setup}
\label{sec:setup}

We work in 1+1 dimensions and consider a sine-Gordon field $\varphi(t,x)$ coupled to a massive scalar field $\psi(t,x)$ according to the Lagrangian density
\beq
\mathcal{L}=\frac{1}{2}\dot{\varphi}^2-\frac{1}{2}\varphi'^2 - \frac{m^2}{\kappa^2}(1-\cos
(\kappa\varphi))+\frac{1}{2}\dot{\psi}^2-\frac{1}{2}\psi'^2 -\frac{1}{2}\mu^2\psi^2-\frac{\lambda}{2} (1-\cos(\kappa\varphi)) \psi^2\,.
\eeq
Here $m$, $\mu$ are the masses of the two fields, $\kappa$ is a parameter introduced for future convenience and $\lambda$ is a coupling constant. We will assume that conditions are such that the $\varphi$ field can be treated classically, while the $\psi$ field is treated fully quantum mechanically. In general, the (potentially space and time-dependent) classical ``background'' field $\varphi$ excites the quantum ``radiation'' field $\psi$, and the excitations of $\psi$ in turn backreact on $\varphi$. In the following it will be useful to make the 
field redefinition $\varphi\rightarrow \phi=\kappa\varphi$ and to work with the rescaled field $\phi(t,x)$. With this new field variable, the Lagrangian density becomes
\beq
\mathcal{L}=\frac{1}{\kappa^2}\left[\frac{1}{2}\dot{\phi}^2-\frac{1}{2}\phi'^2 - m^2(1-\cos
\phi)\right]+\frac{1}{2}\dot{\psi}^2-\frac{1}{2}\psi'^2 -\frac{1}{2}\mu^2\psi^2-\frac{\lambda}{2} (1-\cos\phi) \psi^2\,.
\label{fullmodel}
\eeq
We see that the coupling between the two fields $\phi$ and $\psi$ is such that the discrete shift symmetry 
$\phi\to\phi+2\pi n$, $n\in \mathbb{Z}$,
is maintained.

\subsection{Neglecting backreaction}
\label{subsec:nobkrxn}

\subsubsection{Background dynamics}
\label{subsubsec:bkg_dyn}

We start by discussing the limit $\kappa\rightarrow 0$ which corresponds to the case where the dynamics of 
the $\phi$ field doesn't feel the presence of the radiation field $\psi$. 
If $\psi$ is entirely neglected, this is also the limit when $\phi$ is classical because, in the path integral, $\kappa \to 0$ is 
equivalent to $\hbar \to 0$.
In this ``no backreaction'' case, the $\phi$ field equation reduces to the sine-Gordon equation
\beq
\ddot{\phi}-\phi'' +m^2\sin\phi=0\,.
\label{sineGordon}
\eeq
It is worth mentioning that, in this case, one can define the (conserved) energy of the sine-Gordon field unambiguously by
\beq
E_{\phi}=\frac{1}{\kappa^2}\int dx \left[\frac{1}{2}\dot{\phi}^2+\frac{1}{2}\phi'^2 + m^2(1-\cos
\phi)\right]\,,
\eeq
and it is easy to see that 
the $\kappa\rightarrow 0$ limit is equivalent to the energy of the sine-Gordon field $\phi$ being much larger than the  (renormalized) vacuum energy of the field $\psi$.
We will come back to this point in Sec.~\ref{sec:cases}. 

Eq.~\eqref{sineGordon} has well-known kink and anti-kink solutions,
\beq
\label{eq:singlesoliton}
\phi_{\pm}(t,x)=\pm 4 \arctan\left(e^{\gamma m(x-x_0-v(t-t_0))}\right)\,,
\eeq
which describe a soliton (or antisoliton depending on the sign) whose center is at $x_0$ at time $t_0$ and which moves to the right with velocity $v$, with %
$\gamma=1/{\sqrt{1-v^2}}$
being 
the Lorentz factor. It is worth noting that the energy of this configuration is \beq
E_{\pm}(v)=\frac{8\gamma m}{\kappa^2}\, .\eeq

The integrability properties of the sine-Gordon equation allow for the analytical construction of more complicated solutions involving a kink and an antikink. The first one is the so-called {\it breather} solution which can be understood as a bound state of a kink and an antikink and which reads
\beq
\phi_{\rm breather}(t,x)=4\arctan\left(\frac{\eta\sin(\omega t)}{\cosh(\eta\omega x)}\right)\,.
\label{eq:breathershape}
\eeq
Here $\omega$ is the angular frequency of the breather and $\eta={\sqrt{m^2-\omega^2}}/{\omega}$. 
The energy of the breather is
\beq
E_{\rm breather}=\frac{16\eta\omega}{\kappa^2}=\frac{16m}{\kappa^2}\sqrt{1-\frac{\omega^2}{m^2}}\,,
\eeq
and is seen to be less than the sum of the energy of a static kink and a static antikink. 

The other important solution, which will be the main focus of our attention in the remainder of this paper, can be obtained from the breather solution by making the formal substitution $\omega=im\gamma v$. It reads
\beq
\label{eq:phiKKbar}
\phi_{K\bar{K}}(t,x)=4\arctan\left(\frac{\sinh(\gamma mvt)}{v\cosh(\gamma m x)}\right)\,,
\eeq
and describes the elastic scattering of a kink and an antikink infinitely separated at $t=-\infty$ and moving towards each other with velocity $v$. The collision occurs at $t=x=0$ and the kink and antikink pass through
each other with a time delay. 
As expected, the energy of this field configuration is simply the sum of the energies of a kink and an antikink moving with velocity $v$, 
\beq 
\label{eq:kinkantikinkenergy}
E_{K\bar{K}}=\frac{16\gamma m}{\kappa^2} =2 E_{\pm}(v)\, .
\eeq
Having summarized some important features of the sine-Gordon equation, we are  ready to study the quantum field $\psi$ living in the background of the latter solution. 

\subsubsection{Quantum radiation}
\label{subsubsec:qm_rad}

Integrability of the sine-Gordon model implies that the scattering of a kink-antikink pair is trivial i.e. never forms a bound state, and  that a breather never decays. These are classical properties of the sine-Gordon model but they are maintained when taking the quantum fluctuations of the field into account. However, in our case, the classical sine-Gordon field is coupled to an external quantum scalar field and the energy exchange between the classical field configuration and the quantum bath is expected to invalidate these properties.

We thus turn our attention to the quantum radiation that occurs during kink-antikink scattering,
equivalently particle production in the time-dependent background $\phi_{K\bar{K}}(t,x)$ given in Eq.~\eqref{eq:phiKKbar},
using the framework of  the classical- quantum correspondence 
(CQC)~\cite{Vachaspati:2018llo, Vachaspati:2018hcu, Olle:2019skb} that we  briefly describe 
below.
Since in the limit $\kappa\rightarrow 0$ the dynamics of the 
background field $\phi$ is insensitive to the presence of the field $\psi$,
we will work with the truncated Lagrangian density
\beq
{\mathcal{L}}_{\psi}=\frac{1}{2}\dot{\psi}^2-\frac{1}{2}\psi'^2 -\frac{1}{2}\left\{\mu^2+\lambda\left[1-\cos\phi_{K\bar{K}}(t,x)\right]\right\} \psi^2\,.
\eeq
It is easy to understand why this model leads to excitations of the quantum field $\psi$. 
Indeed this Lagrangian density describes a free scalar field with a space and time dependent 
mass-squared $M^2(t,x)=\mu^2+\lambda\left[1-\cos\phi_{K\bar{K}}(t,x)\right]$. The non-adiabatic variation
of $M^2(t,x)$ will lead to particle production and is expected to occur mostly at the time of kink-antikink
collision at $t=0$.

To study the scattering more quantitatively using numerical methods,
we start by compactifying the spatial dimension on a circle of length $L$ which we then discretize 
on a regular lattice consisting of $N$ evenly spaced points. The lattice spacing is therefore $a={L}/{N}$. 
We can further define the discretized field values $\psi_i=\psi(t,-L/2+ia)$ 
and $\phi_i=\phi_{K\bar{K}}(t,-L/2+ia)$ for $i=1,\ldots, N$, and use the following differencing scheme to estimate the 
second spatial derivative,
\be
\psi''(t,ia)\longrightarrow \frac{\psi_{i+1}-2\psi_i+\psi_{i-1}}{a^2}\,.\label{forwardbackwarddiff}
\ee
With these conventions, and after a spatial integration by parts, the Lagrangian of this discretized model 
can  be written as
\beq
{L}_{\psi,\rm disc.}=\frac{a}{2}\dot{\bm{\psi}}^T.\dot{\bm{\psi}}-\frac{a}{2}\bm{\psi}^T.\bm{\Omega}^2.\bm{\psi}\,,
\eeq
where we have arranged the discretized values of the field $\psi$ in a column vector $\bm{\psi}=(\psi_1,\psi_2,\dots,\psi_N)^T$ and introduced the $N\times N$ matrix
\beq
[\bm{\Omega}^2]_{ij} = 
\begin{cases}
+{2}/{a^2}+\mu^2+\lambda\left[1-\cos{\phi}_i\right]\,,& i=j\\
-{1}/{a^2}\,,& i=j\pm1\ (\text{mod}\ N)\\
0\,,&\text{otherwise}\,.
\end{cases}
\label{Omega2}
\eeq
In this form, the discretized Lagrangian is immediately seen to describe a collection of $N$ harmonic oscillators (located at each lattice point) coupled to each other via quadratic interactions. The total energy and energy density at each lattice point $i$ for the $\psi$ field in this discretized model are also well-defined and given by\footnote{Here, in order to get a more accurate discretized estimate of the local gradient energy, the average between its forward differencing and backward differencing approximations is used. This allows us to integrate Eq.~\eqref{energynonintegrated} into Eq.~\eqref{energyintegrated}. Notice however that simply choosing a forward or backward differencing scheme for the estimation of this component of the energy density would also have been consistent.}
\ba
{H}_{\psi,\rm disc.}&=&\frac{a}{2}\dot{\bm{\psi}}^T.\dot{\bm{\psi}}+\frac{a}{2}\bm{\psi}^T.\bm{\Omega}^2.\bm{\psi}\,,\label{energyintegrated}\\
{\mathcal{H}}_{\psi,\text{disc.},i}&=& \frac{1}{2}\dot{\psi}_{i}^2+\frac{1}{4a^2}\left[(\psi_{i+1}-\psi_{i})^2+(\psi_{i}-\psi_{i-1})^2\right]+\frac{1}{2}\left\{\mu^2+\lambda\left[1-\cos{\phi}_i\right]\right\}\psi_{i}^2\,.\label{energynonintegrated}
\ea

The next step is to quantize the theory (in the Heisenberg picture) by promoting the discretized field values $\psi_i$ to operators $\hat{\psi}_i$ and introducing the time-dependent matrix $\bm{Z}$ whose (complex) elements
$Z_{ij}$ satisfy the relation
\beq
\hat{\psi}_i=Z_{ij}^*\hat{a}_j(t_0) +Z_{ij}\hat{a}^\dag_j(t_0)\,.
\eeq
Here the reference time $t_0$ is chosen so that the background for  $t\leq t_0$ can be  approximated by a
slowly-moving kink-antikink pair separated by a large distance (much larger than their typical size).
This is equivalent to specifying a reference vacuum state $|0\rangle$, the state annihilated by all the $\hat{a}_i(t_0)$. (The ladder operators $\hat{a}_i(t_0)$ and $\hat{a}^\dag_i(t_0)$ refer to the quantum harmonic oscillators located at each lattice point.) Now the dynamics of the field $\hat{\psi}$ is simply given by the Heisenberg equations 
\beq
\ddot{\bm{Z}}+\bm{\Omega}^2.\bm{Z}=0\,,
\label{CQCeqs}
\eeq
with initial conditions 
\beq
\bm{Z}(t_0)=-\frac{i}{\sqrt{2a}}\bm{\Omega}(t_0)^{-1/2}\quad\text{and}\quad \dot{\bm{Z}}(t_0)=\frac{1}{\sqrt{2a}}\bm{\Omega}(t_0)^{1/2}\,.
\label{CQCics}
\eeq
Since $\Omega^2$ is a symmetric positive definite matrix, $\Omega^{\pm1/2}$ is computed by first diagonalizing $\Omega^2$ and then applying the desired power function to its positive eigenvalues.
Notice that these initial conditions
only define the vacuum of the theory unambiguously when the background $\bm{\Omega}^2(t)$ is approximately constant for $t\leq t_0$. In other words if the time evolution of the background before the time $t_0$ is adiabatically slow, a different choice of initial time smaller than $t_0$ will not modify the quantum dynamics.
More precisely the initial conditions~\eqref{CQCics} define the $0$-th order adiabatic vacuum (which corresponds to the lowest order approximation of the mode functions within the WKB approximation~\cite{Birrell:1982ix}). In particular, if the background is such that no $t_0$ obeying 
the required properties can be found, then we can expect spurious excitations of the field $\psi$ to occur 
and one should strive to minimize them. In the case at hand, the  kink and antikink are moving towards 
each other with velocity $v$ at $t=-\infty$ and we find ourselves exactly in the latter situation. However, 
we expect that for small $v$ (in practice less than $0.3$) the initial conditions~\eqref{CQCics} will still 
give reasonable results (and we will come back to this later in Sec.~\ref{sec:results}). 

\subsubsection{Observables}

The evolution of $\psi$ is given by the system of equations given in matrix form in Eq.~\eqref{CQCeqs}. Initialized by Eqs.~\eqref{CQCics}, they are particularly simple and they can readily be solved numerically. Moreover, the $Z_{ij}$ variables allow for the easy computation of various observable quantities of interest. For example,  the expression for the vacuum expectation value of the energy in the radiation field can be written as
\be
E_\psi\equiv\langle 0| \hat{H}_{\psi,\rm disc.}|0\rangle=\frac{a}{2}\text{Tr}\left[\dot{\bm{Z}}^\dag.\dot{\bm{Z}}+\bm{Z}^\dag.\bm{\Omega}^2.\bm{Z}\right]\,.
\ee
This is equal to the total (classical) energy in the $Z_{ij}$ variables. Similarly, the vacuum expectation value of the energy density at the $i$-th lattice point is
\begin{align}
\rho_{\psi,i} & \equiv\langle 0| \hat{\mathcal{H}}_{\psi,\text{disc.},i}|0\rangle\,\nonumber\\ & = \sum_{j=1}^N\left( \frac{1}{2}|\dot{Z}_{ij}|^2+\frac{1}{4a^2}\left[|Z_{i+1,j}-Z_{ij}|^2+|Z_{ij}-Z_{i-1,j}|^2\right]+\frac{1}{2}\left\{\mu^2+\lambda\left[1-\cos{\phi}_i\right]\right\}|Z_{ij}|^2\right)\,.
\end{align}
 Moreover the spatial two-point function of the radiation field can be written as
\be
C_{ij}\equiv\langle 0| \hat{\psi}_i\hat{\psi}_j |0\rangle = \sum_{k=1}^N Z^*_{ik}Z_{jk}\,.
\label{twopoint}
\ee
In order to render these expressions insensitive to the discretization scale $a$, we first renormalize the spatial two-point function by subtracting its $\lambda=0$ counterpart,
\be
C_{ij}^{(R)}=C_{ij}-\left.C_{ij}\right|_{\lambda=0}\,,
\label{correnorm}
\ee
and then use it to renormalize the $\lambda$-dependent part of the energy and energy density. This procedure is closely related to point-splitting~\cite{Birrell:1982ix}. The resulting renormalized quantities are 
thus
\ba
\rho_{\psi,i}^{(R)}&=&\rho_{\psi,i}-\frac{1}{2}\lambda\left[1-\cos\phi_i\right] \left.C_{ii}\right|_{\lambda=0}-\left.\rho_{\psi,i}\right|_{\lambda=0}\,,\label{renormenden}\\
E_{\psi}^{(R)}&=&E_\psi-\sum_{i=1}^{N}\frac{a}{2}\lambda\left[1-\cos\phi_i\right] \left.C_{ii}\right|_{\lambda=0}-\left.E_\psi \right|_{\lambda=0}\,.\label{renormen}
\ea
In both these equations, the last term corresponds to the subtraction of the (constant) zero-point energy.

We can therefore compute any quantity of interest for the study of $\psi$ particle production in the $\phi_{K\bar{K}}$ background by studying the classical dynamics of the variables $Z_{ij}$ (with well-chosen initial conditions). We thus trade $N$ real quantum variables (corresponding to the discretized field values $\psi_i$) for $N\times N$ complex classical variables (the $Z_{ij}$). This is the essence of the CQC and it is closely related to the mode function and Bogoliubov coeffcient methods \cite{Vachaspati:2018hcu}.

\subsubsection{Initial conditions and vacuum structure}

\begin{figure}[htb]
\begin{center}
\includegraphics[width=0.65\textwidth,angle=0]{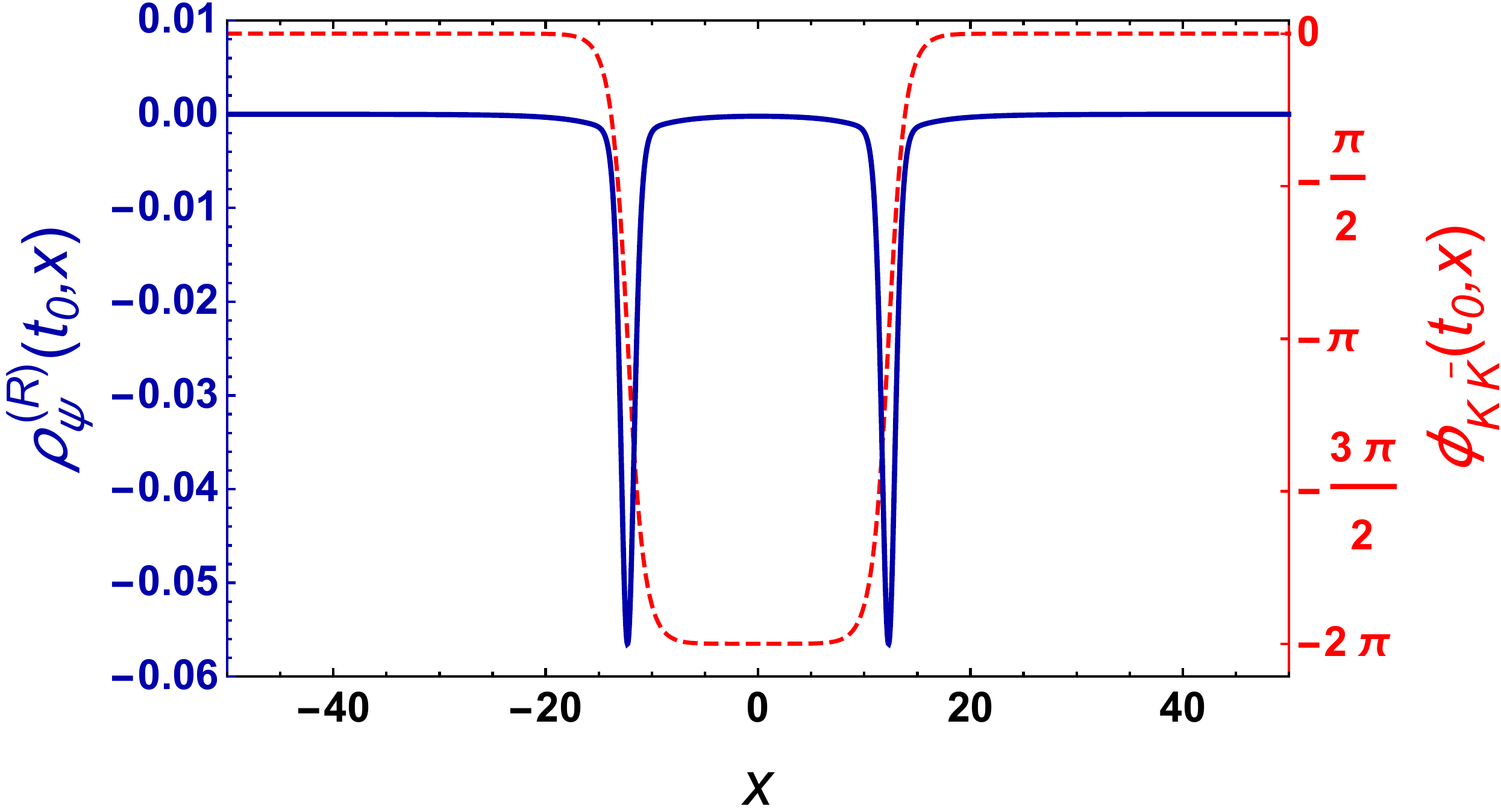}
\caption{\label{fig:psi_cloud_eg} Example of the renormalized initial energy density in $\psi$ and the initial kink-antikink profile. 
The kinks are dressed in $\psi$ particles whose energy within the kink is lower than that outside the kink,
{\it i.e.} they form bound states.
The  kink-antikink profile is shown as a red dashed line (vertical scale on the right) and the clouds of $\psi$ particles (represented by the renormalized energy density of $\psi$) are shown in dark blue (vertical scale on the left). The parameters are $L=100$, $N=500$, $m=1$, $v = 0.1$, $\mu = 0.1$, $\lambda=0.3$, $\kappa \rightarrow 0$ and the initial time is chosen as $t_0 = -100$. 
}
\end{center}
\end{figure}

\begin{figure*}
    \begin{center}
         \subfloat[] {\includegraphics[width=0.45\textwidth]{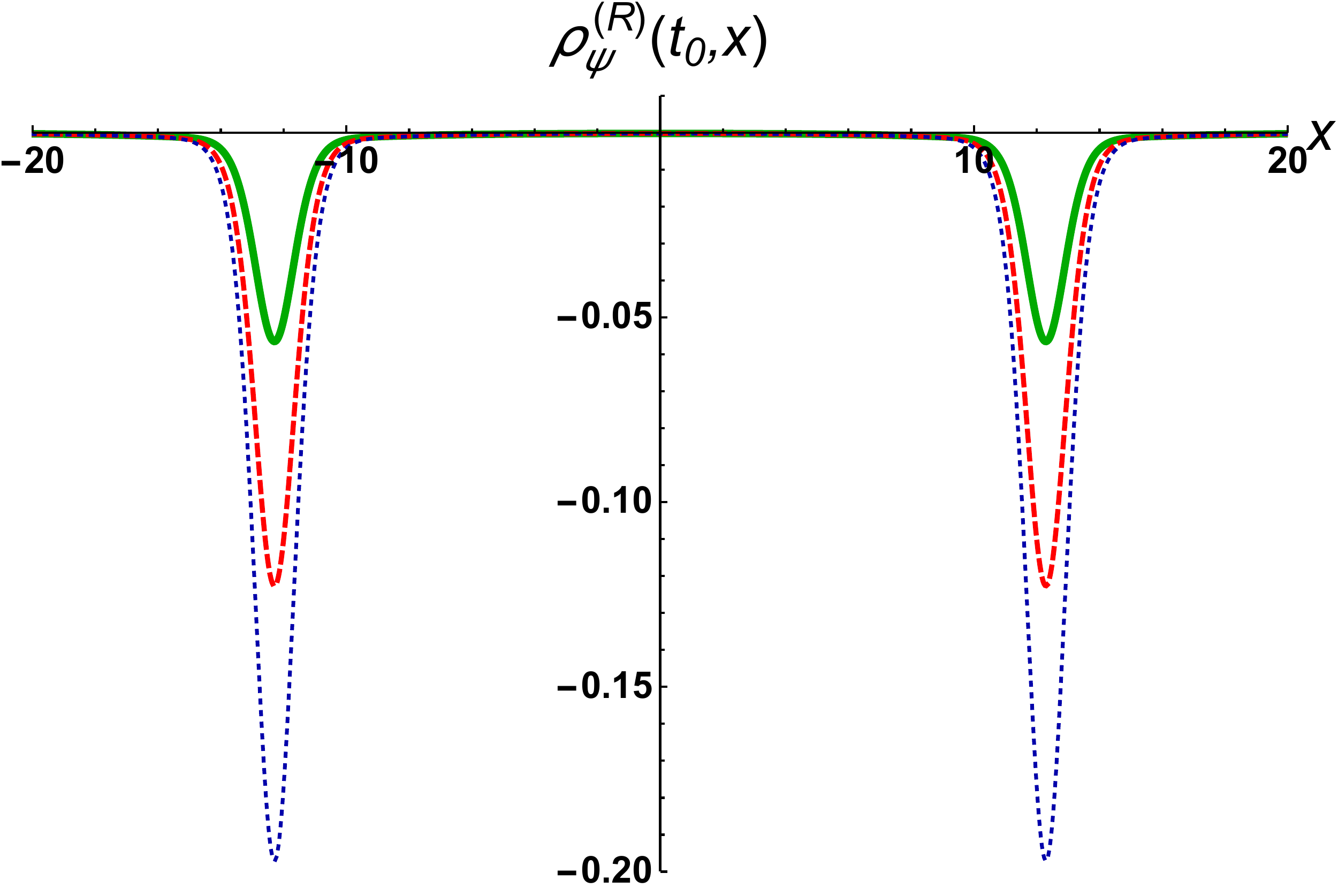}\label{subfig:lamdepic}}\hfill%
        \subfloat[] {\includegraphics[width=0.45\textwidth]{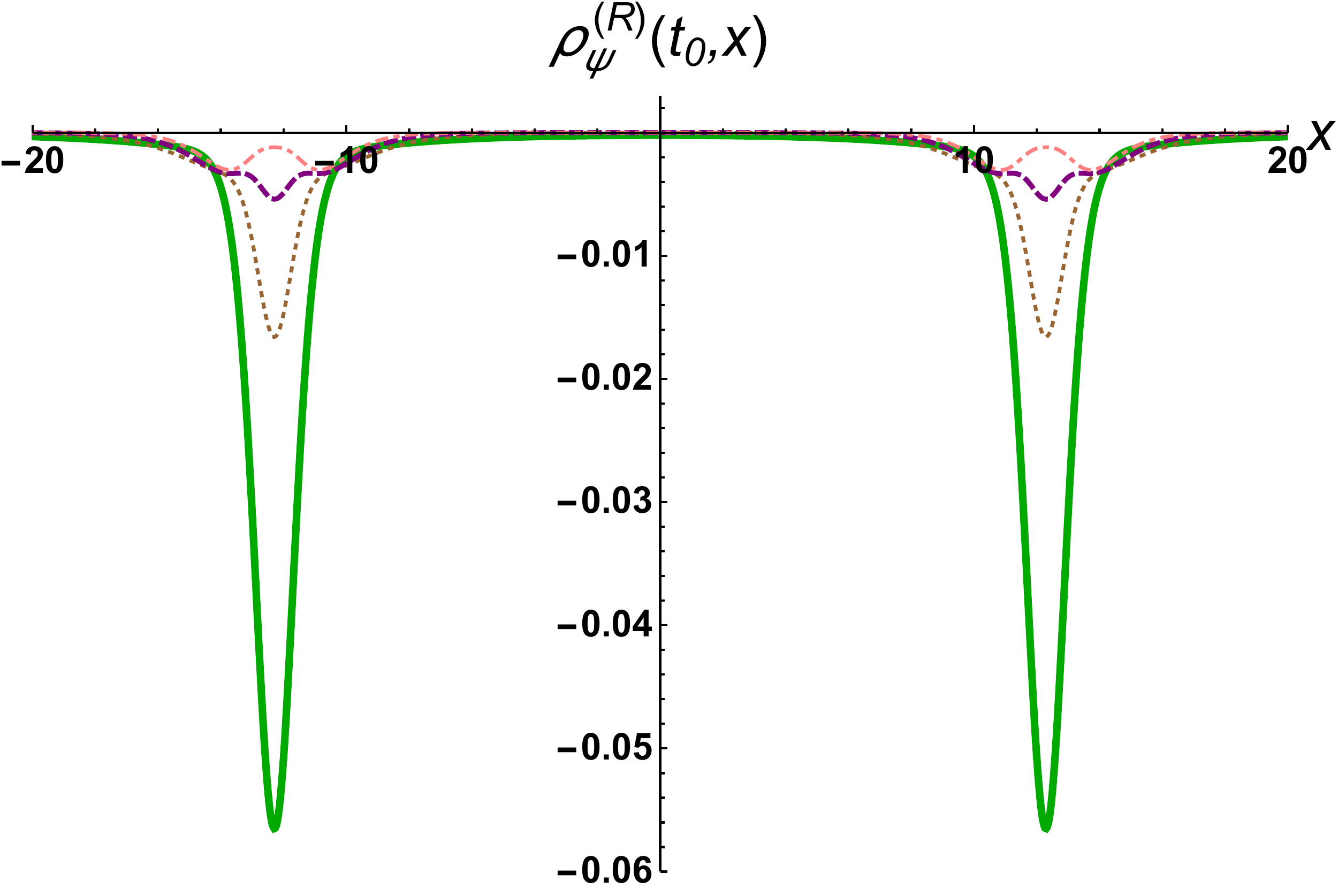}\label{subfig:mudepic}}\hfill
        \caption{(a) Renormalized initial energy density in $\psi$ ($\rho^{(R)}_\psi$) for various values of $\lambda$: 0.3 (solid green), 0.5 (dashed red), 0.7 (dotted dark blue) for $\mu = 0.1$. (b) Renormalized initial energy density in $\psi$ for various values of $\mu$: 0.1 (solid green), 0.3 (dotted brown), 0.5 (dashed purple), 0.7 (dot-dashed pink) for $\lambda = 0.3$. The other parameters are $L=100$, $N=500$, $m=1$, $v = 0.1$ , $\kappa \rightarrow 0$ and the  initial time is $t_0 = -100$. Note that the solid green lines correspond to the same set of parameters ($\lambda=0.3$, $\mu=0.1$) in both the panels.
        }
        \label{fig:enden_ic}
    \end{center}
\end{figure*}

The renormalized energy density observable 
$\rho_{\psi,i}^{(R)}$
allows us to visualize the vacuum structure of the quantum field $\psi$.  Fig.~\ref{fig:psi_cloud_eg} shows the renormalized energy density in the $\psi$ field at the initial time superimposed over the background kink-antikink profile. (The parameter $a$ will always be chosen such that the lattice provides a good approximation of the continuum limit and therefore we will liberally identify the discretized quantities $\rho^{(R)}_{\psi, i}(t)$ and $\phi_i(t)$  to their  continuous counterparts $\rho^{(R)}_{\psi}(t,x)$ and $\phi(t,x)$ for $x=-L/2+ia$.) We see that the presence of the background induces a dip
in the energy density  around the kink and antikink, akin to two clouds of $\psi$ particles.
The width of these clouds is set by the kink and antikink width i.e. $1/m$. However, as shown in Fig.~\ref{fig:enden_ic}, the depth of the underdensities
depends on the coupling $\lambda$ and the mass of the $\psi$ field $\mu$.
In particular, the trough
of the $\psi$ energy density around the kink-antikink position increases for larger $\lambda$ and smaller $\mu$. Moreover, for some parameters, the energy density of $\psi$ shows some non-trivial features that reflect the relevance of the additional length scales $\mu^{-1}$ and $\lambda^{-1/2}$.

\subsection{Including backreaction}
\label{subsec:bkrxn}

The results above assumed that backreaction can be neglected. This is certainly a good approximation when $\kappa$ is small but what happens when backreaction can no longer be neglected? 
This constitutes the main part of this work, to which we now turn.

We thus consider the case $\kappa\neq 0$ so that the field equation for the $\phi$ field gets a contribution from the $\psi$ field:
\be
\ddot{\phi}-\phi'' +\left(m^2+\frac{\lambda\kappa^2}{2}\psi^2\right)\sin\phi=0\,.
\ee
Of course this equation is fully classical and one would need to decide how the quantum excitations in the $\psi$ 
field couple to the classical background $\phi$ by modifying the coupling term proportional to $\psi^2\sin\phi$. 
The most straightforward way is to use the 
semiclassical approximation\footnote{
Another procedure for incorporating backreaction is based on the stochastic method, where the initial conditions for the field $\psi$ are sampled from a distribution similar to Eq.~\eqref{CQCics}. Several simulations, each with different samples for the initial conditions, would be performed and then averaged. We leave the interesting comparison between the two methods for future work.} which simply entails replacing $\psi(t,x)^2$ by $\langle 0|\hat{\psi}(t,x)^2|0\rangle$. 
Therefore the $\phi$ field 
equation becomes
\be
\ddot{\phi}-\phi'' +\left(m^2+\frac{\lambda\kappa^2}{2}\langle 0|\hat{\psi}^2|0\rangle\right)\sin\phi=0\,.
\ee

To make contact with the methods described in the previous section, we need the discretized version of this equation,
\be
\ddot{\phi}_i-\frac{1}{a^2}(\phi_{i+1}-2\phi_i+\phi_{i-1})+\left(m^2+\frac{\lambda\kappa^2}{2}\langle 0|\hat{\psi}_i^2|0\rangle\right)\sin\phi_i=0\,,
\label{semiclassicalapprox}
\ee
where $\phi_i=\phi(t,-L/2+ia)$ is not given by the non-dynamical kink-antikink solution $\phi_{K\bar{K}}$ anymore, 
rather it will have to be solved for subject to the initial conditions $\phi_i(t_0)=\phi_{K\bar{K}}(t_0,-L/2+ia)$ and 
$\dot{\phi}_i(t_0)=\dot{\phi}_{K\bar{K}}(t_0,-L/2+ia)$. Solving Eq.~\eqref{semiclassicalapprox} requires being able 
to compute the dynamics of the quantity $\langle 0|\hat{\psi}_i^2|0\rangle$. This can be done by using the results 
of the previous section and in particular Eq.~\eqref{twopoint} to make the substitution
\be
\langle 0|\hat{\psi}_i^2|0\rangle=\sum_{j=1}^N|Z_{ij}|^2
\ee
with $Z_{ij}$  given by Eqs.~\eqref{CQCeqs} and~\eqref{CQCics}, where background field values ${\phi}_i$
in the definition of $\bm{\Omega}^2$ are replaced by the corresponding dynamical values. 
Because $\sum_{j=1}^N |Z_{ij}|^2$ is a 1+1 dimensional two-point function in the coincident point 
limit, in other words $C_{ii}$ given in Eq.~\eqref{twopoint}, it is logarithmically sensitive to the discretization scale $a$, and it would 
produce infinite backreaction in the continuum limit~\cite{Vachaspati:2018llo, Vachaspati:2018hcu, Olle:2019skb}.
This can be remedied by noticing that the parameter $m$ appearing in the field equations is actually the bare 
mass of the sine-Gordon field, and using it to renormalize the divergence. We therefore define the physical 
mass $m_{\rm phys}$ of the sine-Gordon field by
\be
\label{eq:massrenorm}
m^2=m^2_{\rm phys}-\frac{\lambda\kappa^2}{2}\sum_{j=1}^N \left. |Z_{ij}|^2 \right|_{\lambda=0}
=m^2_{\rm phys}-\frac{\lambda\kappa^2}{4a}\left[\bm{\Omega}_0^{-1}\right]_{ii}
=m^2_{\rm phys}-\frac{\lambda\kappa^2}{4L}{\rm Tr}\,\bm{\Omega}_0^{-1}\,.
\ee
Here, in the second equality, we used the fact that, when $\lambda=0$, 
$\bm{\Omega}^2(t)\equiv\bm{\Omega}^2_0$ is a constant matrix, and therefore 
Eq.~\eqref{CQCeqs} has the simple solution
\be
\bm{Z}(t)=-\frac{i}{\sqrt{2a}}e^{i\bm{\Omega}_0 (t-t_0)}\bm{\Omega}_0^{-1/2}\,.
\ee
Moreover in the third equality of Eq.~\eqref{eq:massrenorm} we use the fact that all the diagonal coefficients 
of the $\bm{\Omega}_0^{-1}$ 
matrix are equal to each other (see Eq.~\eqref{Omega2}). Overall this procedure is equivalent to replacing the two-point function 
$C_{ii}$ by its renormalized counterpart $C_{ii}^{(R)}$ in Eq.~\eqref{semiclassicalapprox}, and $m$ by 
$m_{\rm phys}$ in every equation where it appears. 
Notice that, in the non-backreacting case ($\kappa\rightarrow 0$), the bare mass $m$ and the physical 
mass $m_{\rm phys}$ are one and the same. Henceforth, we will simply choose $\kappa=1$ when 
taking backreaction into account.
 
 Summing up, the backreacted dynamics (within the semiclassical approximation) are given by the system of coupled differential equations 
\begin{equation}
\ddot{\phi}_i - \frac{1}{a^2}(\phi_{i+1}-2\phi_i+\phi_{i-1}) + \left[m_{\rm phys}^2 
+ \frac{\lambda\kappa^2}{2}\sum_{j=1}^N \left(|Z_{ij}|^2-\left. |Z_{ij}|^2 
\right|_{\lambda=0}\right)\right]\sin\phi_i = 0\,, 
\label{backreactedphieq}
\end{equation}
and
\begin{equation}
\ddot{\bm{Z}}+\bm{\Omega}^2\bm{Z}=0 \,,  
\label{backreactedZeq}
\end{equation}
with initial conditions 
\ba
\phi_i(t_0)=\phi_{K\bar{K}}(t_0,-L/2+ia)\,,\quad&&\quad \dot{\phi}_i(t_0)=\dot{\phi}_{K\bar{K}}(t_0,-L/2+ia)\,,\label{backreactedphiICs}\\
\bm{Z}(t_0)=-\frac{i}{\sqrt{2a}}\bm{\Omega}(t_0)^{-1/2}\,,\quad&&\quad \dot{\bm{Z}}(t_0)=\frac{1}{\sqrt{2a}}\bm{\Omega}(t_0)^{1/2}\,,
\label{backreactedZICs}
\ea
where we recall that 
\be
\phi_{K\bar{K}}(t,x)=4\arctan\left(\frac{\sinh(\gamma m_{\rm phys}vt)}{v\cosh(\gamma m_{\rm phys} x)}\right)
\ee
and where the matrix $\bm{\Omega}^2$  defined in Eq.~\eqref{Omega2} depends on the fully dynamical background 
$\phi_i(t)$.
Note that the mass of the $\phi$ field used in the initial conditions is the physical mass $m_{\rm phys}$, as shown in the above equation.

Before discussing the results of our simulations, let us say a few words about the observables that we are going to use in our study of particle production and the associated backreaction during kink-antikink collisions. These need to accurately account for the energy exchanged between the background and the quantum radiation bath. Of course, in general, in a fully interacting theory it is impossible to separate the energy of a particular subsystem. However, as long as the coupling $\lambda$ is not too big (smaller than $m_{\rm phys}^2$ in practice) we can assume that Eqs.~\eqref{renormen} and~\eqref{renormenden} still provide a useful measure of the (renormalized) energy, respectively  energy density, in the quantum field $\psi$. Under these assumptions, the energy in the kink-antikink background is given by
\be
E_{\phi}\equiv \frac{a}{\kappa^2}\sum_{i=1}^N \left[\frac{1}{2}\dot{\phi}_i^2+\frac{1}{4a^2}\left[(\phi_{i+1}-\phi_{i})^2+(\phi_{i}-\phi_{i-1})^2\right] + m_{\rm phys}^2(1-\cos
\phi_i)\right]\,.
\label{Ephi}
\ee
Notice that the total (conserved) energy of the coupled system is given by 
\be
\label{eqn:totalen}
E=E_{\phi}+E_{\psi}^{(R)}\,,
\ee
where $E_\phi$ and $E_\psi^{(R)}$ are given by Eq.~\eqref{Ephi}  and Eq.~\eqref{renormen} respectively. 

\section{Results}
\label{sec:results}

In this section we present our numerical results. 
We will always work in units where $m_{\rm phys}=1$.
which is equivalent to rescaling space, time, $\mu$ and $\lambda$ by $m_{\rm phys}$.
Our lattice is a circle of physical size $L=100$ and is sampled by $N=500$ equally spaced points ($a=0.2$). We use an explicit Crank-Nicholson method with two iterations
to solve the system of coupled ordinary differential equations given in Eqs.~\eqref{backreactedphieq} and~\eqref{backreactedZeq} with the initial conditions of Eqs.~\eqref{backreactedphiICs} and~\eqref{backreactedZICs}.
The initial time  is taken to be $t_0=-100$. Because most interesting effects are expected to occur at or around the time of collision, $t=0$, and to make sure that finite lattice size effects don't spoil any physical effects, we only evolve the equations for one light-crossing time after the collision i.e. up to $t=L=100$. We have also made sure that the results do not strongly depend on the UV and IR cutoffs ($a=L/N$ and $L$) and that our renormalization scheme is sufficient to eliminate any UV sensitivity (see Fig.~\ref{appfig:ndep_ldep_check}).
Of course, decreasing $a$ would improve the resolution of the simulations but, given the $N^2$ complexity 
of the code, this is not feasible without significantly more computational cost.
Similarly a larger lattice would allow us to track the dynamics of the system for a longer time but 
maintaining spatial resolution would again require a corresponding increase in $N$. In the current work, energy non-conservation due to 
numerical error {\it over the whole duration of time evolution} is of the order of $0.1\%$.

The range over which we will vary the different parameters $\lambda$, $v$ and $\mu$ will depend on 
numerical accuracy as well as on the intrinsic limitations of our model. To avoid strong coupling effects 
we will be allowing $\lambda$ to vary between 0 and 1. 
Since our initial conditions for the 
$Z_{ij}$ variables are strictly only valid when the background is adiabatically varying, we limit ourselves 
to velocities $v$ smaller than 0.3.\footnote{Because of this violation of adiabaticity at $t=t_0$, a small amount of spurious particle production is observed for the $v=0.3$ case but the corresponding dissipated energy is of the order of $0.1\%$ of $E_\phi$ which can safely be neglected.} Moreover, we require the kink-antikink pair to be well-separated at the 
initial time $t_0$ which, because of the finite size of our periodic lattice, limits the range of $v$ 
to the interval $(0.05, 0.4)$.
Since the background field $\phi$ is assumed to behave classically, its mass 
$m_{\rm phys}$ is expected to be larger than the mass of the radiation field $\psi$. Thus $\mu$ is constrained 
to be smaller than 1. 

We start by presenting the two qualitatively different possible outcomes of the inelastic kink-antikink scattering studied in this paper: pure scattering, or formation of a bound state, {\it i.e.}
a breather-like structure. Then we will discuss what role the different parameters of the model play in the 
occurrence of these two outcomes.

\subsection{Scattering or formation of a bound state}

\begin{figure}
    \begin{center}
         \subfloat[] {\includegraphics[width=0.45\textwidth]{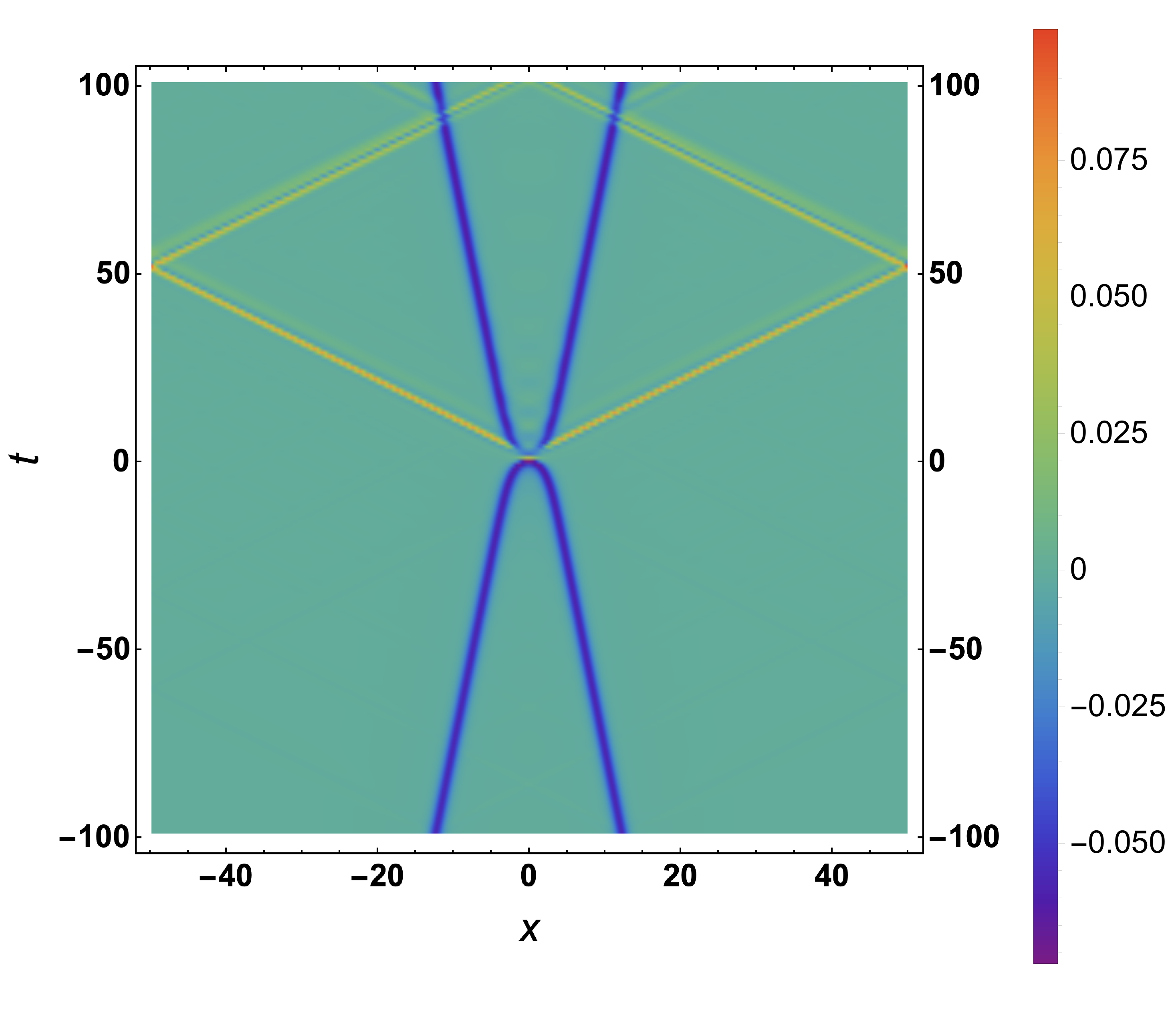}\label{subfig:denplot1a}}\hfill%
        \subfloat[] {\includegraphics[width=0.45\textwidth]{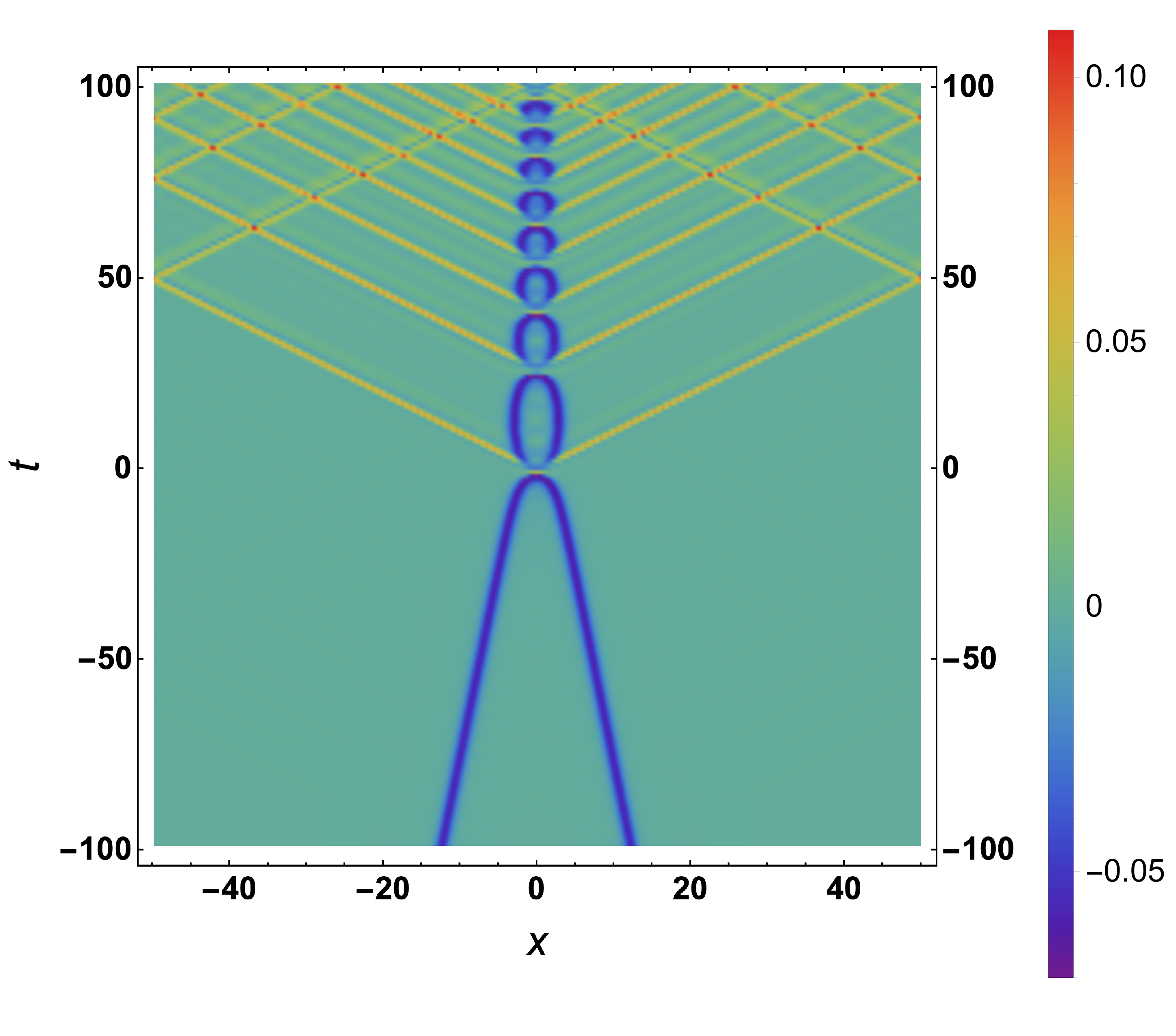}\label{subfig:denplot1b}}\hfill
        \\
        \subfloat[] {\includegraphics[width=0.45\textwidth]{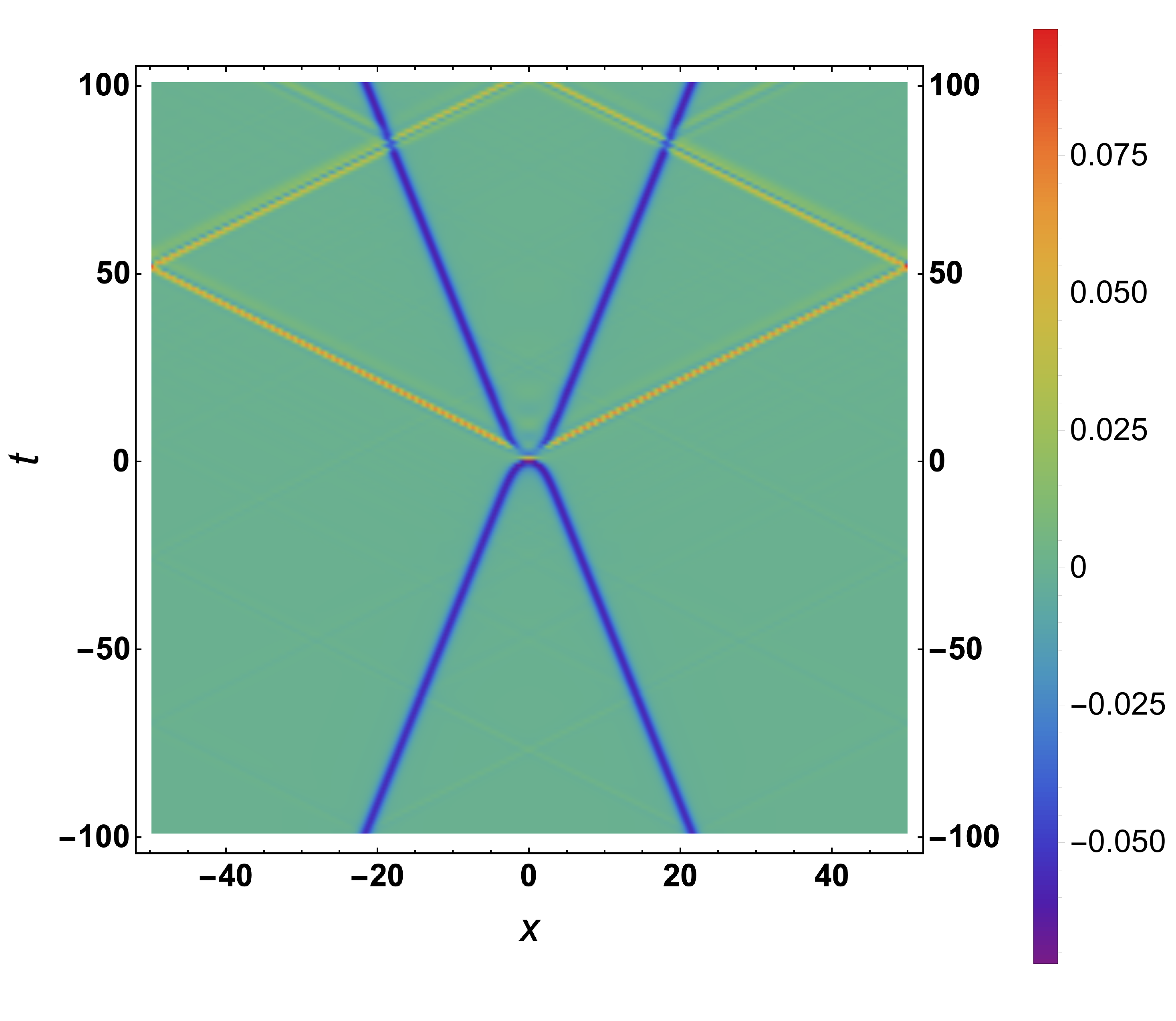}\label{subfig:denplot2a}}\hfill%
        \subfloat[] {\includegraphics[width=0.45\textwidth]{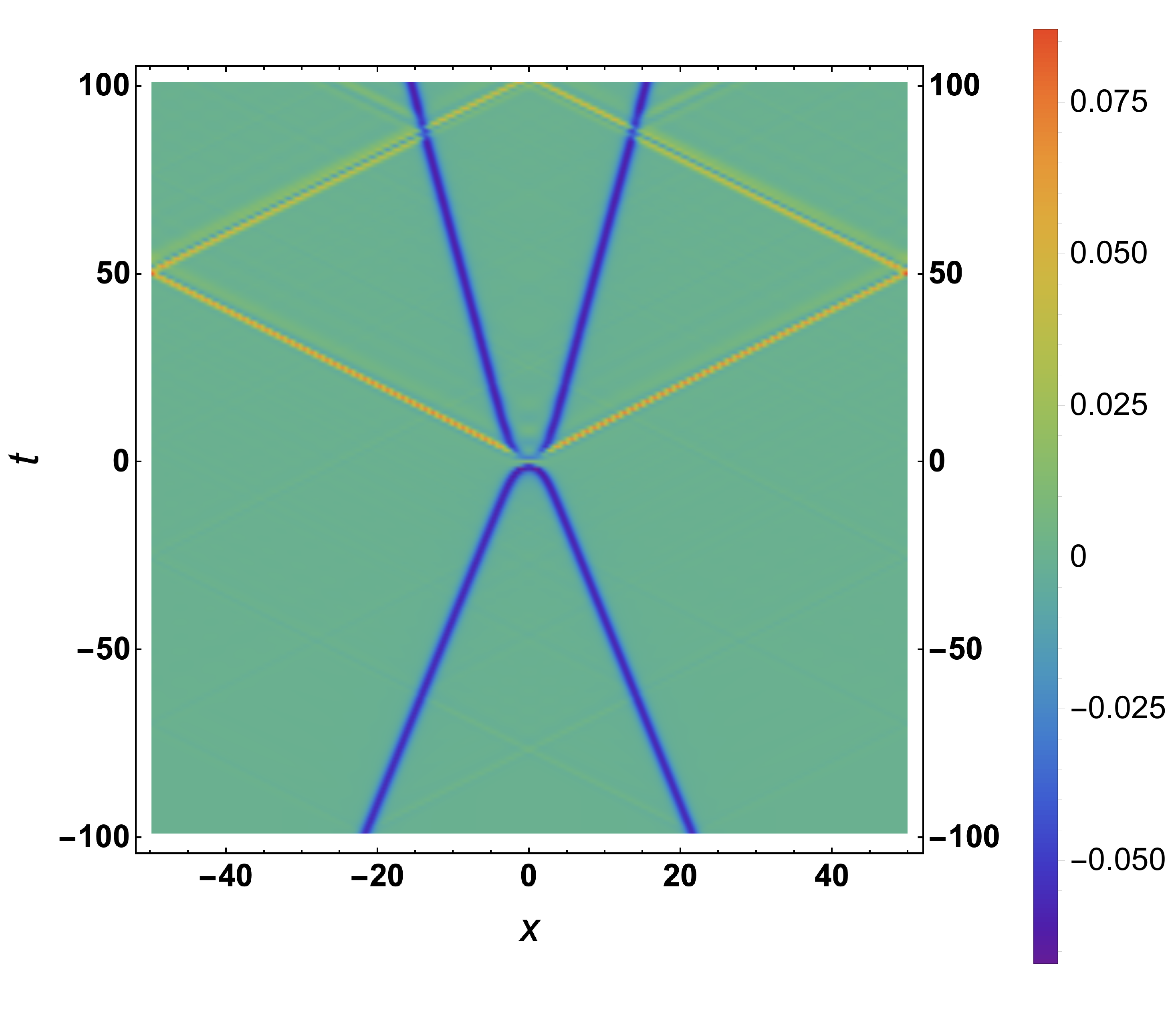}\label{subfig:denplot2b}}\hfill
        \caption{Time evolution of the renormalized energy density in $\psi$ ($\rho^{(R)}_\psi$). 
        (a) For, $\kappa \rightarrow 0$, i.e., no backreaction; $v = 0.1$, $\lambda=0.3$; 
        (b) For, $\kappa = 1$, i.e., with backreaction; $v =0.1$, $\lambda=0.3$. 
        (c) For, $\kappa \rightarrow 0$, i.e., no backreaction; $v = 0.2$, $\lambda=0.3$; 
        (d) For, $\kappa = 1$, i.e., with backreaction; $v =0.2$, $\lambda=0.3$. The universal parameters 
        are $L=100$, $N=500$, $\mu = 0.1$, $m_{\rm phys}=1$, and $t_0 = -100$. The collision happens at $t=0$. The animations corresponding to the different cases can be found at \href{https://sites.google.com/asu.edu/mainakm/animations\#h.oaaca9ogoc42}{https://sites.google.com/asu.edu/mainakm}. 
           }
        \label{fig:density_plots}
    \end{center}
\end{figure}

In Fig.~\ref{fig:density_plots} we show the renormalized energy density in the $\psi$ field as a function of $x$ (horizontal axis) and $t$ (vertical axis) with and without backreaction for two different sets of parameters. The color coding represents the value of the energy density from its minimum value in dark blue to its maximum value in bright red.
Since the clouds of $\psi$ particles (thick dark blue lines) 
track the kink and antikink perfectly, this representation is particularly well suited for visualizing both the dynamics of the kink-antikink background and any radiation bursts (thin orange lines)
occurring during their collision. It serves as an accurate spacetime diagram of the collision. The first thing we notice, in Figs.~\ref{subfig:denplot1a} and~\ref{subfig:denplot2a}, is that when $\kappa\rightarrow 0$ i.e. when backreaction is neglected, the kink and antikink dynamics are unperturbed by the collision (as they should be); there are however two bursts of radiation originating and $x=t=0$ i.e. at the collision, and propagating away from the kink-antikink pair (one to the left the other to the right) at or near the speed of light. When reaching the end of the lattice they wrap around and come back towards the center by virtue of the periodic boundary conditions. The origin of the bursts of radiation can be explained intuitively by the high degree of non-adiabaticity of the background at the moment of collision. 

More interestingly, Figs.~\ref{subfig:denplot1b} and ~\ref{subfig:denplot2b}, including backreaction effects, show two radically distinct behaviors. In Fig.~\ref{subfig:denplot2b}, the parameters are such that the collision does not lead to the formation of a bound state. Just like in the non-backreacting case of Fig.~\ref{subfig:denplot2a}, the  clouds of $\psi$ particles tracking the kink-antikink pair describe an X-like pattern: they are seen to converge, collide and subsequently diverge from one another. The collision is again accompanied by the emission of two bursts of radiation. However, unlike in Fig.~\ref{subfig:denplot2a}, the outgoing  relative velocity of the kink-antikink pair 
in Fig.~\ref{subfig:denplot2b}
is smaller than the initial velocity as background kinetic energy has been converted in quantum radiative energy. In Fig.~\ref{subfig:denplot1b}, the parameters are such that the collision leads to the formation of a breather-like kink-antikink bound state. The cactus (or caterpillar) pattern described by the clouds of $\psi$ particles after the initial collision at $t=0$ represents the multiple subsequent collisions that the kink and antikink undergo, each accompanied by a burst of radiative energy. The first burst of energy depletes enough kinetic energy that the  kink and antikink are not able to break away from their mutual attraction and must form a bound state. We will study the evolution of the energy in $\phi$, $E_{\phi}$, in more detail in the next subsection.

\subsection{Parameter dependence of the outcome of the collision}

\begin{figure}[htb]
\begin{center}
\includegraphics[width=0.65\textwidth,angle=0]{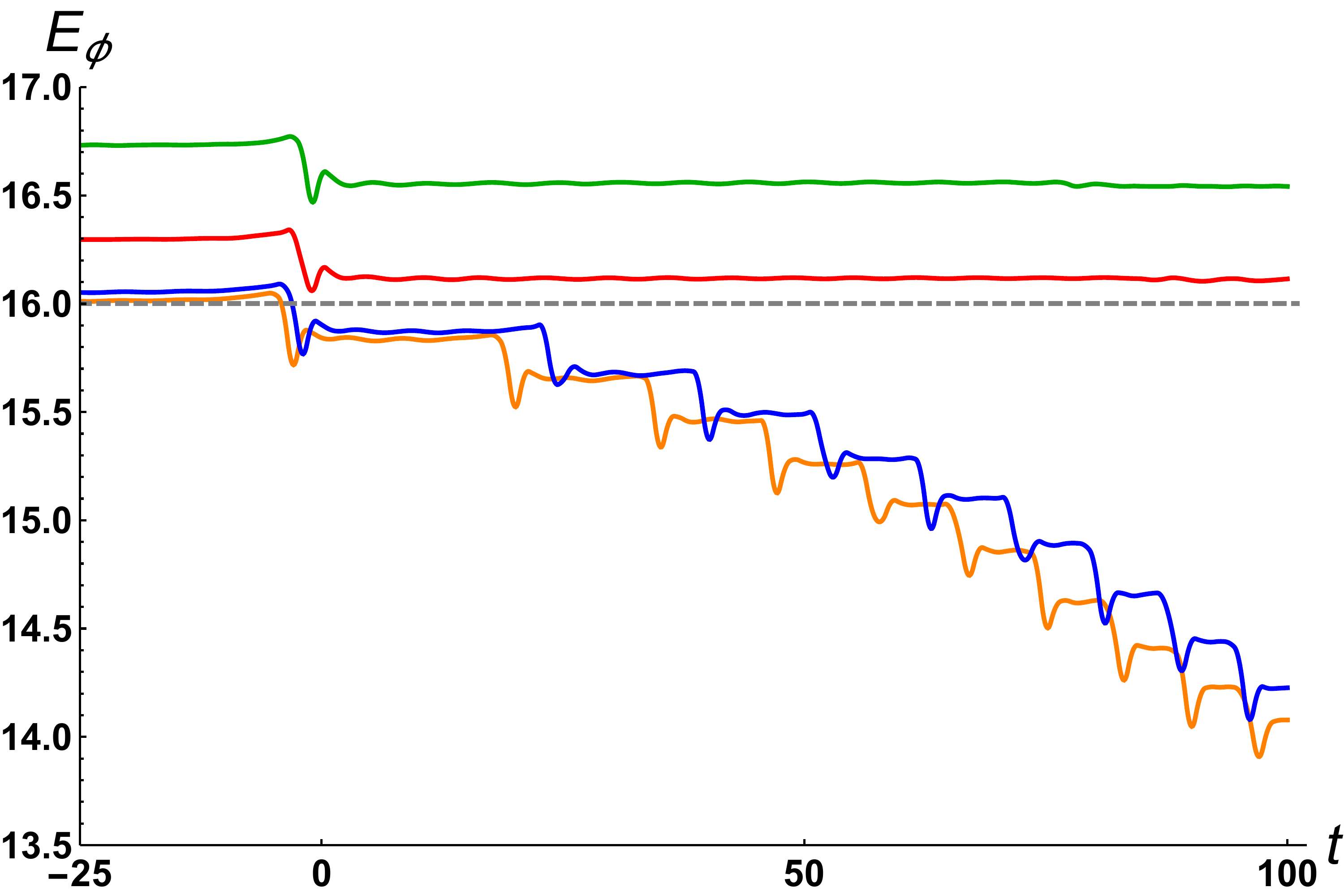}
\caption{\label{fig:vcomp_bkrxn} The energy in the background $\phi$ ($E_\phi$) as a function of time for various values of $v$: 0.07 (Orange), 0.1 (Blue), 0.2 (Red) and 0.3 (Green). The other parameters are $L=100$, $N=500$, $\mu = 0.1$, $m_{\rm phys}=1$, $\lambda = 0.3$ , $\kappa = 1$ and $t_0 = -100$. The gray dashed line corresponds to $E_\phi = 16$.
}
\end{center}
\end{figure}

\begin{figure}[htb]
\begin{center}
\includegraphics[width=0.65\textwidth,angle=0]{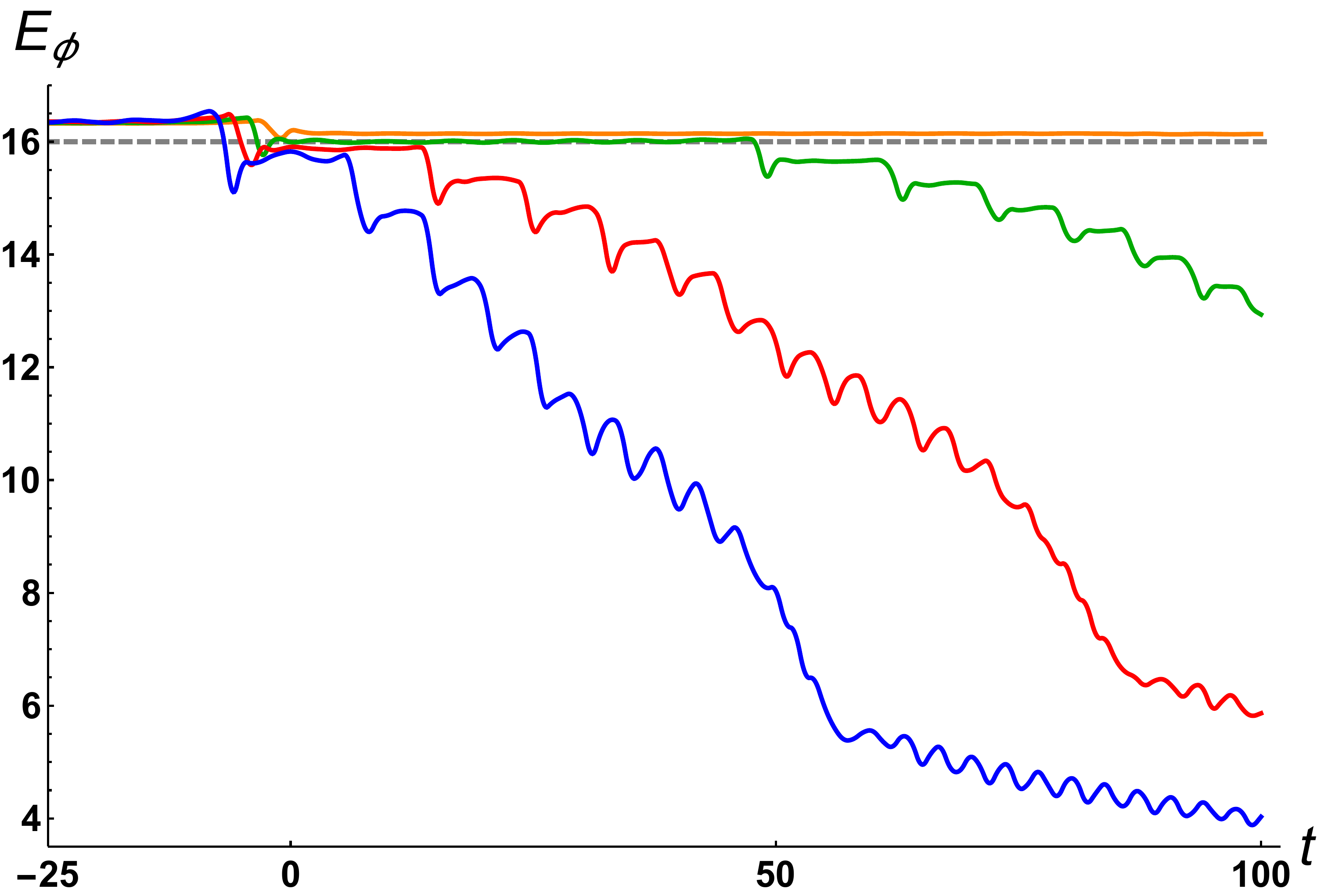}
\caption{\label{fig:lcomp_bkrxn} The energy in the background $\phi$ ($E_\phi$) as a function of time for various values of $\lambda$: 0.3 (Orange), 0.5 (Green), 0.7 (Red) and 0.9 (Blue). The other parameters are $L=100$, $N=500$, $\mu = 0.1$, $m_{\rm phys}=1$, $v = 0.2$ , $\kappa = 1$ and $t_0 = -100$. The gray dashed line corresponds to $E_\phi = 16$.
}
\end{center}
\end{figure}

\begin{figure}[htb]
\begin{center}
\includegraphics[width=0.65\textwidth,angle=0]{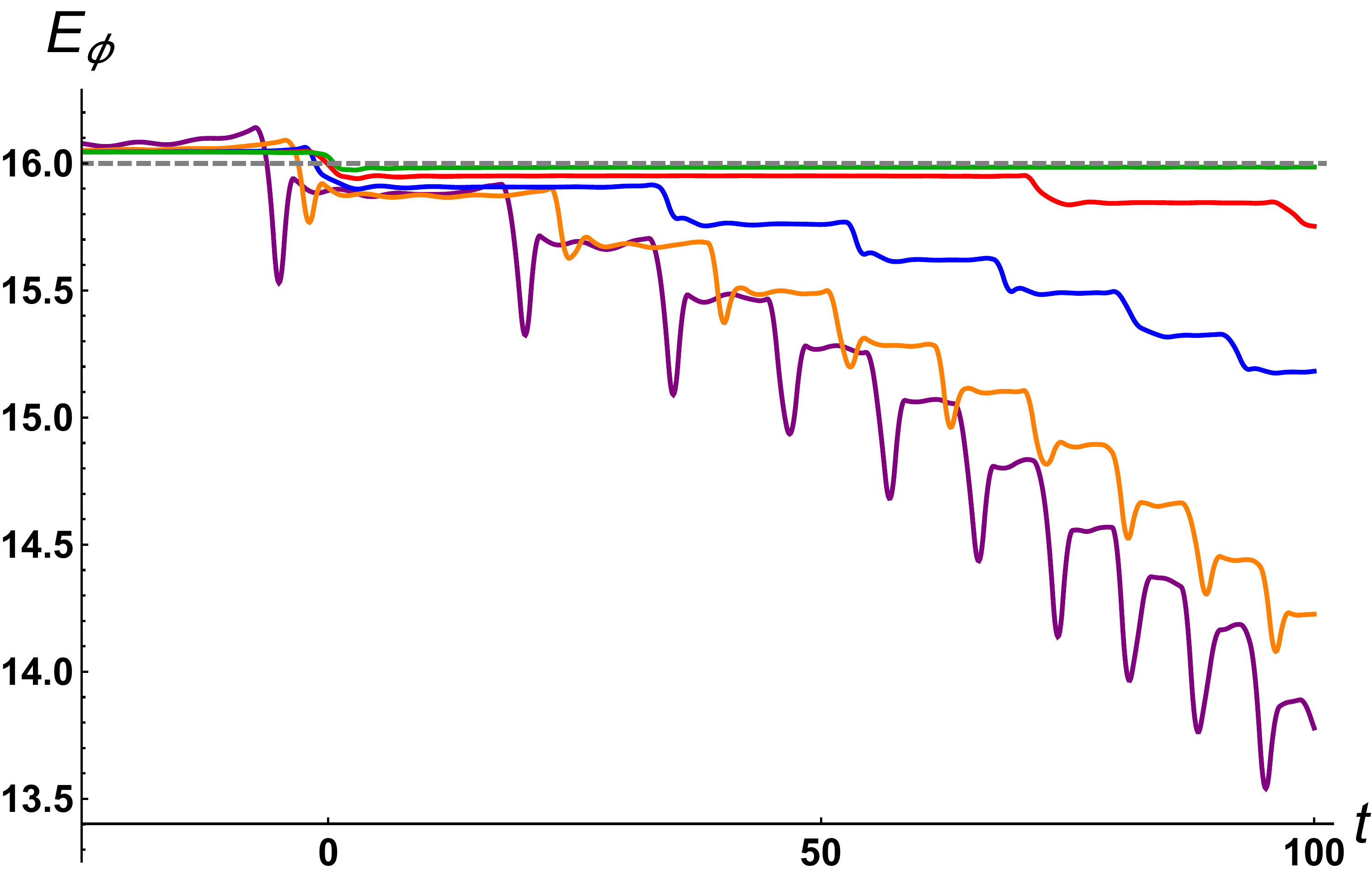}
\caption{\label{fig:mpsicomp_bkrxn} The energy in the background $\phi$ ($E_\phi$) as a function time for various values of $\mu$: 0.01(Purple), 0.1 (Orange), 0.3 (Blue), 0.5 (Red) and 0.7 (Green). The other parameters are $L=100$, $N=500$, $\lambda = 0.3$, $m_{\rm phys}=1$, $v = 0.1$ , $\kappa = 1$ and $t_0 = -100$. The gray dashed line corresponds $E_\phi = 16$.}
\end{center}
\end{figure}

To understand the influence of the parameters of the model on the dynamics of the collision (with backreaction 
taken into account), we turn our attention to the energy in the kink-antikink pair. 
In Figs.~\ref{fig:vcomp_bkrxn},~\ref{fig:lcomp_bkrxn} and~\ref{fig:mpsicomp_bkrxn} we plot $E_{\phi}$ as a 
function of time for different values of the parameters $v$, $\lambda$ and $\mu$. We can make a few 
general comments before going into the details of the plots. First we notice that before the collision $E_{\phi}$ 
is constant, which is to be expected because the kink and antikink are effectively decoupled until their 
relative distance becomes of the order of the kink width $1/m_{\rm phys}$. At the collision, the energy 
decreases abruptly which corresponds to the initial burst of radiation seen in the previous subsection. 
The energy then quickly stabilizes to a new plateau. If the value of this new energy plateau is larger than 
the energy of a static kink-antikink pair i.e. $2E_{\pm}(0)=16m_{\rm phys}$, 
then a breather-like object does not form and the kink-antikink pair remains unbound after the collision 
($E_{\phi}$ remains constant after the collision). On the contrary, if the first burst of energy is large enough  to 
make the value of the energy plateau fall below this threshold energy, then 
a bound state is formed. This is followed by a cascade of subsequent bursts of energy that lead to 
the decay of $E_\phi$ into lower and lower plateaus of shorter and shorter duration. This cascade is readily understood. After the first collision, the kinks separate out more than their widths which leads to the formation of an energy plateau (because kinks radiate only when they overlap). Eventually though, after a certain number of subsequent collisions, they don't manage to separate out and the plateau shape is lost. The radiation thus transitions through
different behaviors which we will discuss in more detail in the next section.

Figure~\ref{fig:vcomp_bkrxn} shows $E_\phi$ for $\lambda=0.3$, $\mu=0.1$, and varying $v$. The higher the initial relative velocity $v$ the bigger the gap to the threshold energy of bound state formation, $\Delta_{\rm gap}\equiv 16(\gamma-1)m_{\rm phys}$, as seen from Eq.~\eqref{eq:kinkantikinkenergy}. We notice that the energy of the first radiation burst doesn't depend strongly on $v$ (see Appendix~\ref{appsec:spect}) and therefore, as long as it is larger than $\Delta_{\rm gap}$, 
a breather-like object is formed. This happens in particular for $v=0.2$ and $v=0.3$.

Figure~\ref{fig:lcomp_bkrxn} shows $E_\phi$ for $v=0.2$, $\mu=0.1$, and varying $\lambda$. 
Here, the energy of the first radiation burst increases with $\lambda$ (it approximately scales as $\lambda^{1.6}$ for the range of parameters considered here) and exceeds $\Delta_{\rm gap}$ for $\lambda=0.5$ and $\lambda=0.7$, when a breather-like object forms. In the $\lambda=0.7$ and $\lambda=0.9$ cases we notice a peculiar change of behavior in $E_\phi$ around $t=70$ when the cascading decay becomes an oscillatory decay with a smaller average slope. The bound state then appears to settle to a long-lived, weakly radiating, oscillon-like object. We will come back to this 
intriguing configuration  in the next section.

Figure~\ref{fig:mpsicomp_bkrxn} represents $E_\phi$ for $v=0.1$, $\lambda=0.3$, and varying $\mu$. Here we notice that the lighter the $\psi$ field, the stronger the initial radiation burst after collision (it scales as $\exp(-1.8\mu)$). This is readily understood, as a light field is more easily excited by the time-dependent background. For $\mu\leq 0.5$, a breather-like object forms since the kink-antikink pair releases an energy greater than $\Delta_{\rm gap}$ at the collision.

\begin{figure}
    \begin{center}
         \subfloat[] {\includegraphics[width=0.45\textwidth]{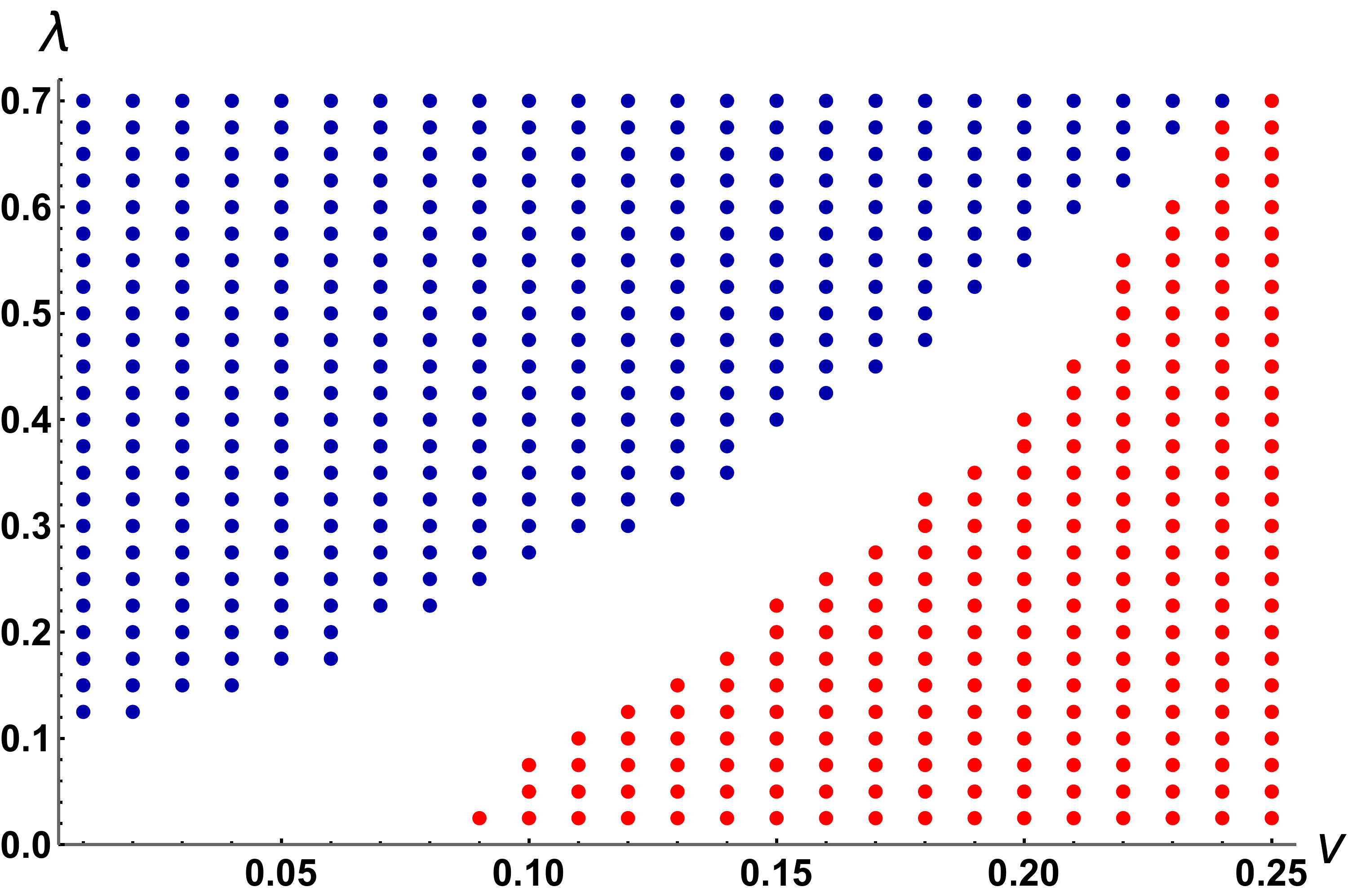}\label{subfig:lvscmp1}}\hfill%
        \subfloat[] {\includegraphics[width=0.45\textwidth]{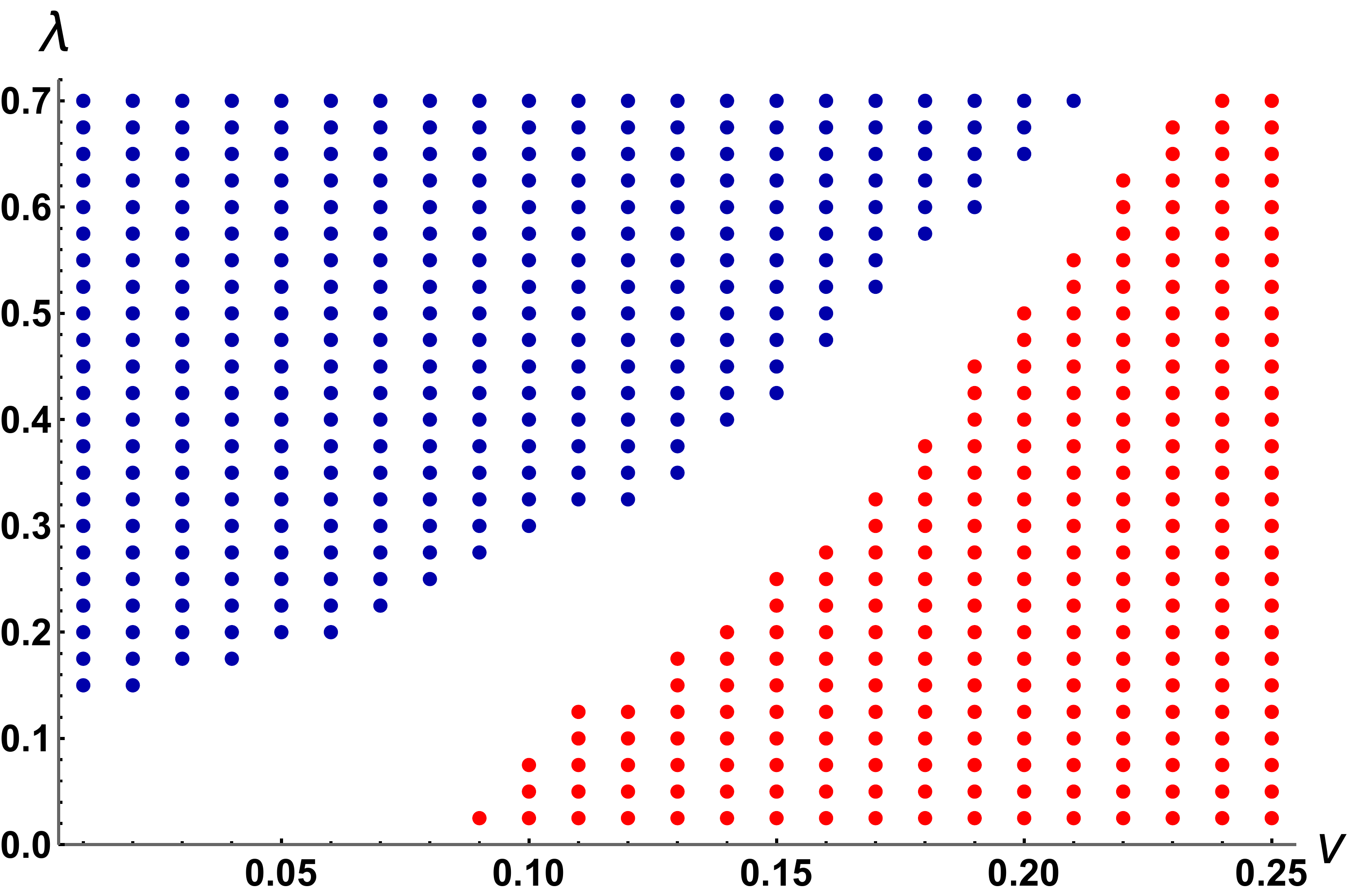}\label{subfig:lvscmp3}}
        \\
        \subfloat[] {\includegraphics[width=0.45\textwidth]{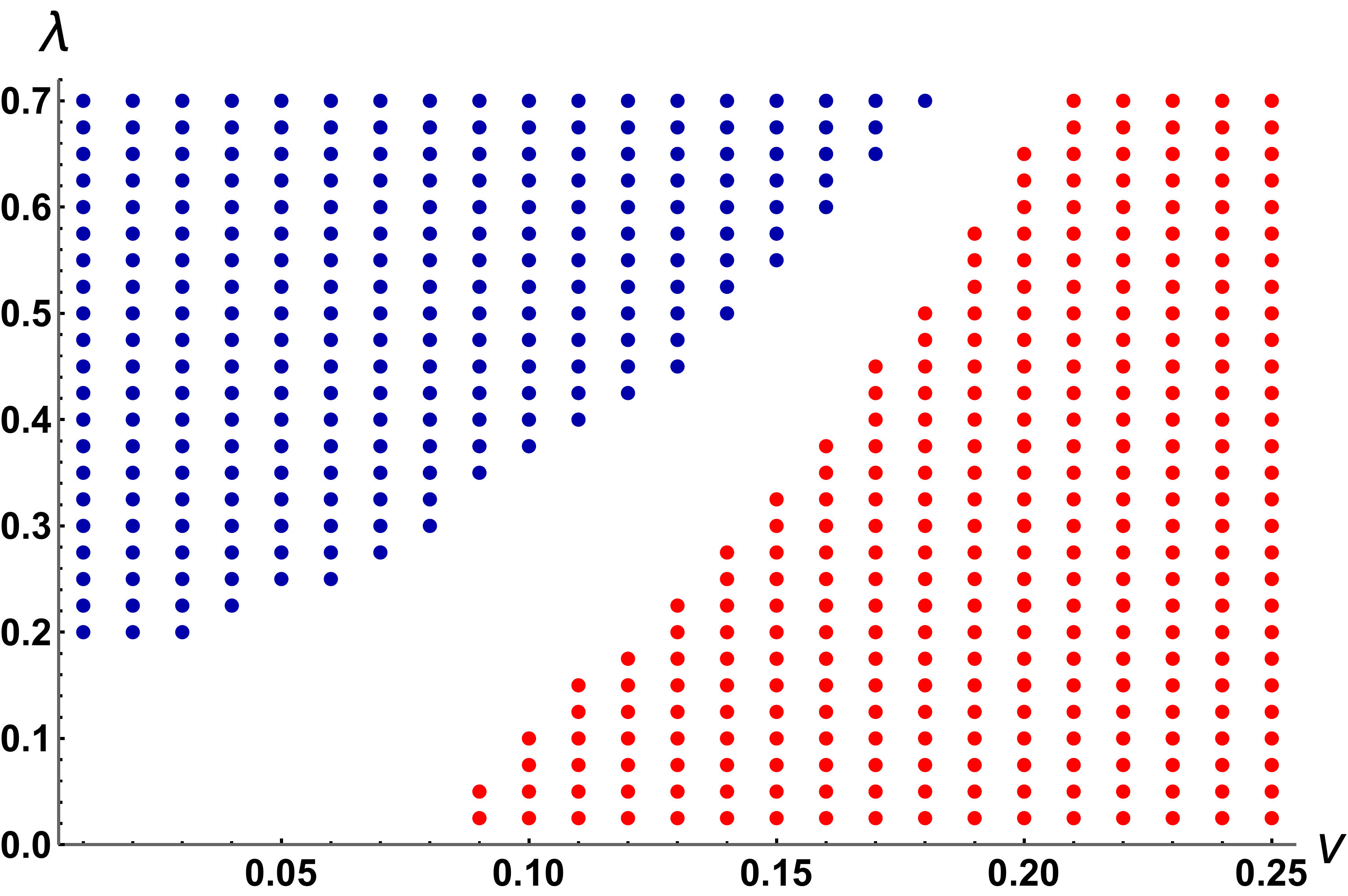}\label{subfig:lvscmp5}}\hfill%
        \subfloat[] {\includegraphics[width=0.45\textwidth]{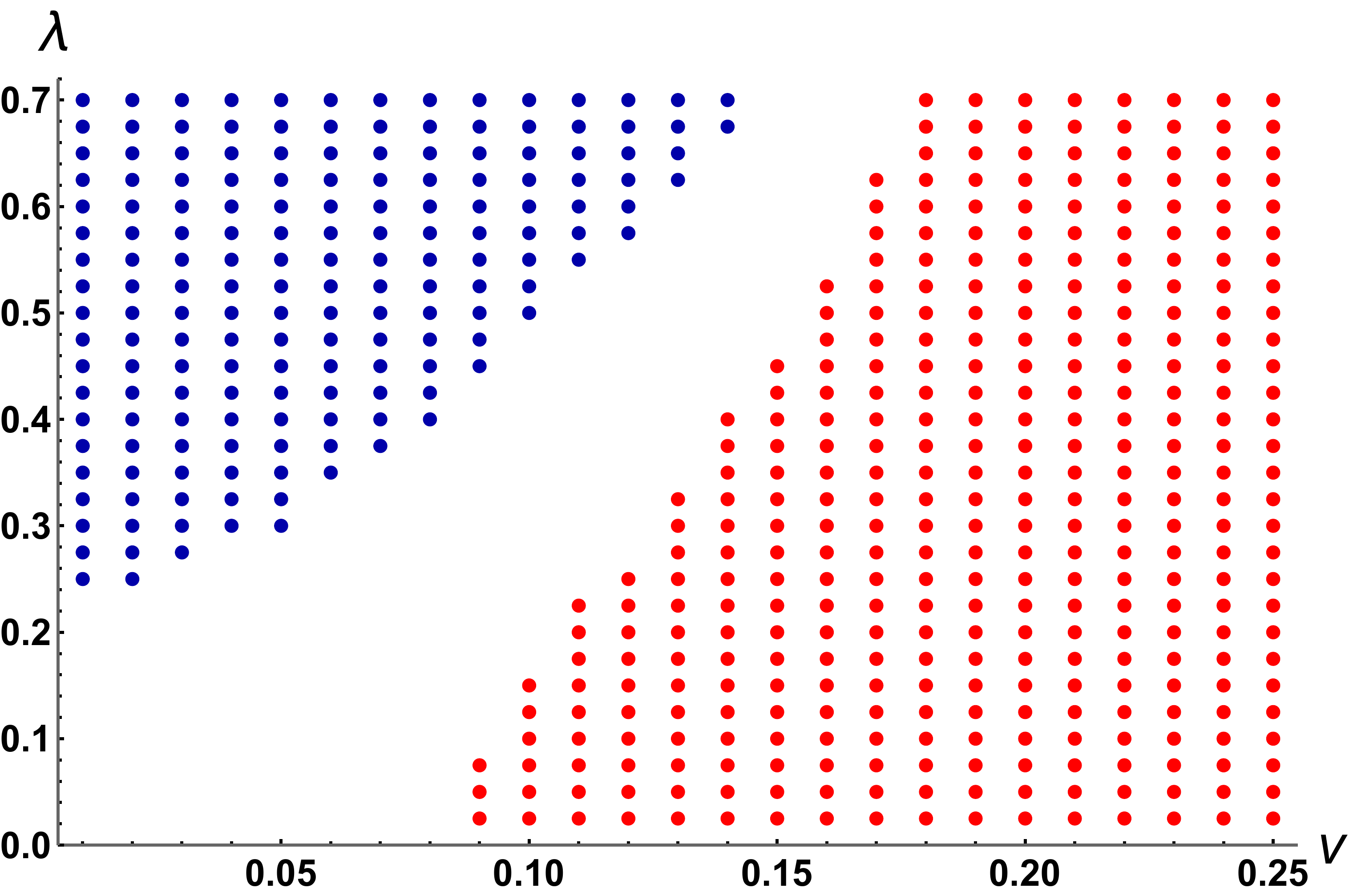}\label{subfig:lvscmp7}}
        \caption{Plots showing a parameter scan of the $\lambda - v$ plane for different values of $\mu$: (a) $\mu = 0.1$, (b) $\mu = 0.3$, (c) $\mu = 0.5$, and (d) $\mu = 0.7$. The dark blue points represent a breather-like object being formed. The blank space denotes the region of uncertainty where a breather-like object may or may not form. The red points denote where a breather-like object does not form. The other parameters are $L=100$, $N=500$, $\kappa = 1$, and $m_{\rm phys}=1$. Because of the long computation time required to generate these plots, we exceptionally choose $t_0=-50$.
        }
        \label{fig:lvscan}
    \end{center}
\end{figure}

We can even go further and determine the region of $(\lambda, v, \mu)$ parameter space where a breather-like object forms, i.e. the region where the first burst of energy at the collision is larger than $\Delta_{\rm gap}$. We expect that the boundary between the region where a bound state forms and the one where the kink-antikink pair remains unbound after collision, to approximately look like a two-dimensional surface. However, since our determination of $\Delta_{\rm gap}$ is imperfect, particularly because the first energy plateau after collision exhibits small oscillations (as can be seen for instance in Figs.~\ref{fig:vcomp_bkrxn},~\ref{fig:lcomp_bkrxn} and~\ref{fig:mpsicomp_bkrxn}), the boundary will necessarily have a thickness. To estimate this thickness we determine the characteristic amplitude of the small oscillations around the first energy plateau in the worst case scenario (by computing the standard deviation from the mean value $\sigma_{\rm max}$ in this case) and we conservatively declare that a first energy burst equal to $\Delta_{\rm gap}$ with a margin of error of $\pm 2\sigma_{\rm max}$ doesn't allow us to definitively determine whether a bound state forms or not. Here $\sigma_{\rm max}=0.027$.
In Fig.~\ref{fig:lvscan} we show the results of a parameter scan with resolution of $0.25$ in $\lambda$, $0.01$ in $v$ and $0.2$ in $\mu$.

\section{Decay of the bound state}
\label{sec:cases}

We have already seen that, when the parameters of the problem are such that a bound-state forms as a consequence of the kink-antikink collision, the radiation goes through different phases. In this section we highlight two of those phases: the fast decay phase where radiation is emitted via successive energy bursts, and the oscillon phase which is weakly radiating and quasi-stable.

\subsection{The energy plateau phase: decay of the breather-like object}
\label{subsec:decayofthebreather}

\begin{figure}
\begin{center}
    \includegraphics[width=0.65\textwidth]{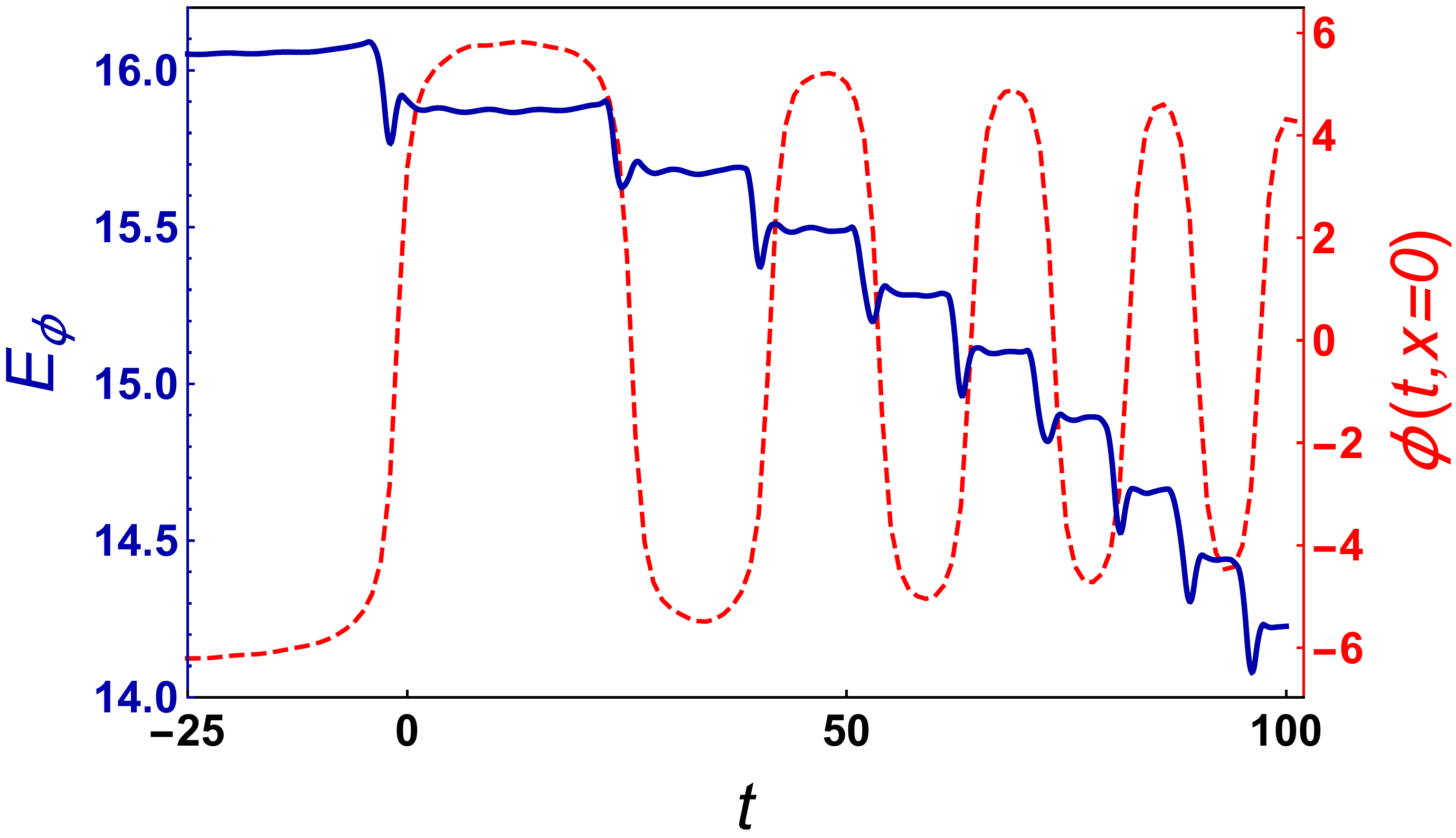}
    \caption{
    Plot of the energy in $\phi$ ($E_\phi$) as a function of time (solid dark blue) superimposed over a plot of $\phi (t,0)$ (dashed-red color). The parameters correspond to the solid blue curve in Fig.~\ref{fig:vcomp_bkrxn}. They are $L=100$, $N=500$, $v=0.1$, $\mu = 0.1$, $\lambda=0.3$, $m_{\rm phys}=1$, $\kappa=1$ and $t_0=-100$.
    }
    \label{fig:plateau_n_vp1}
\end{center}
\end{figure}

\begin{figure*}
    \begin{center}
         \subfloat[] {\includegraphics[width=0.45\textwidth]{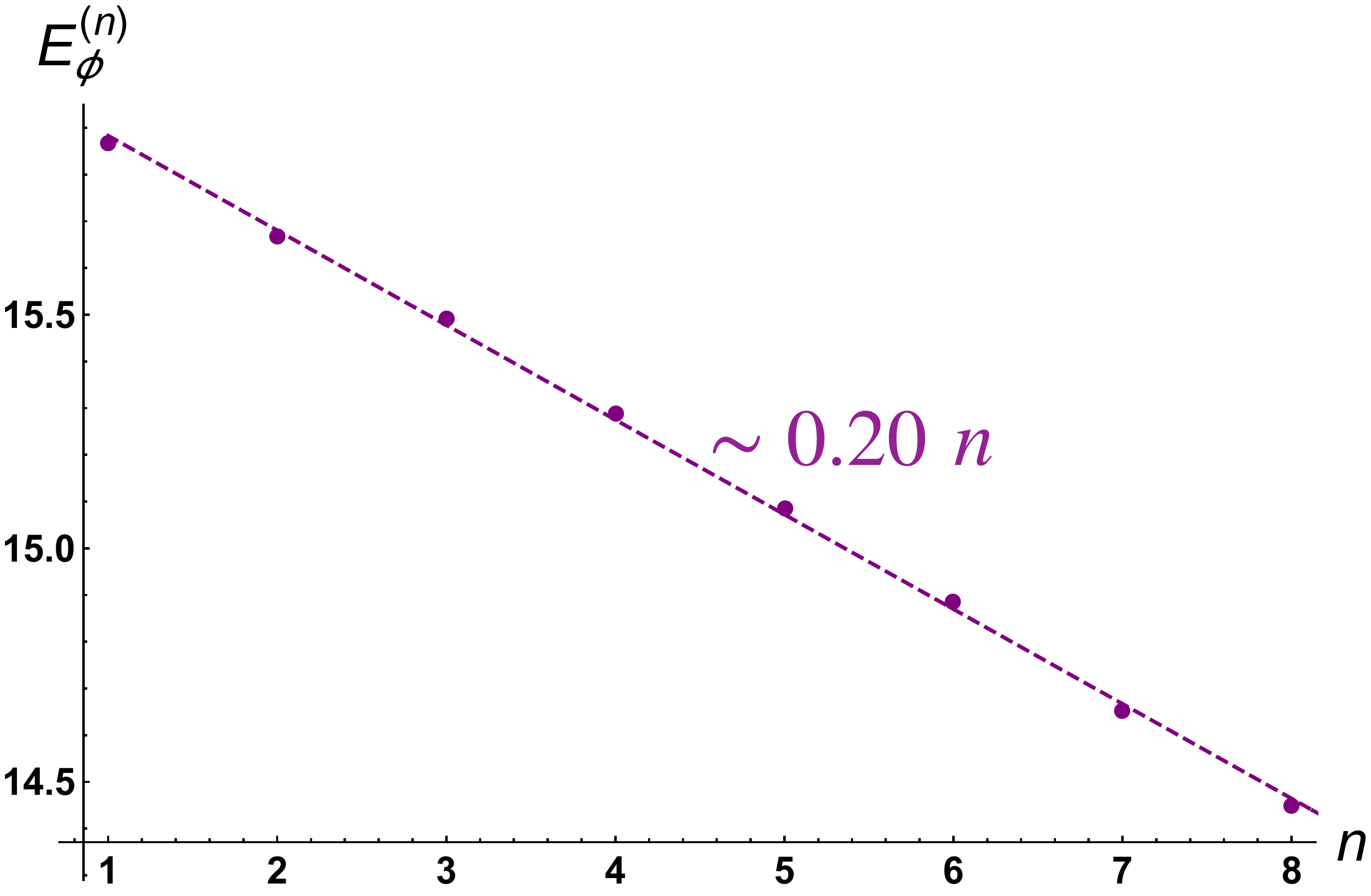}\label{subfig:delEn}}\hfill%
        \subfloat[] {\includegraphics[width=0.45\textwidth]{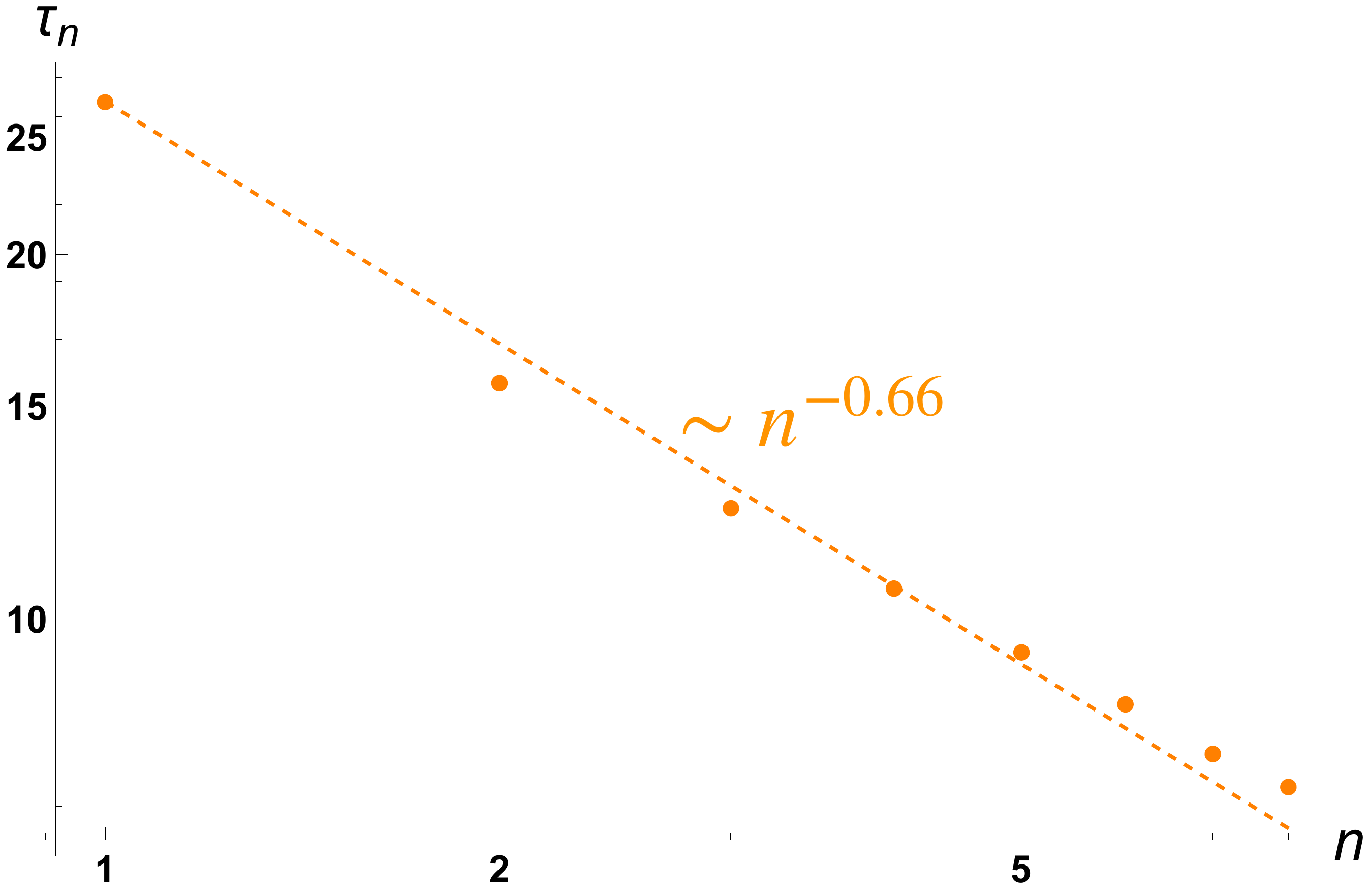}\label{subfig:taun}}\hfill
        \caption{(a) Plot of the energy of the $n$-th plateau of $E_\phi$ in Fig.~\ref{fig:plateau_n_vp1} ($E_{\phi}^{(n)}$) as a function of $n$ (purple); (b) Plot of the duration of said plateau ($\tau_n$) as a function $n$ (orange). The parameters
    are $L=100$, $N=500$, $v=0.1$, $\mu = 0.1$, $\lambda=0.3$, $m_{\rm phys}=1$, $\kappa=1$ and $t_0=-100$. 
        }
        \label{fig:delEtau_n}
    \end{center}
\end{figure*}

\begin{figure}
\begin{center}
    \includegraphics[width=0.65\textwidth]{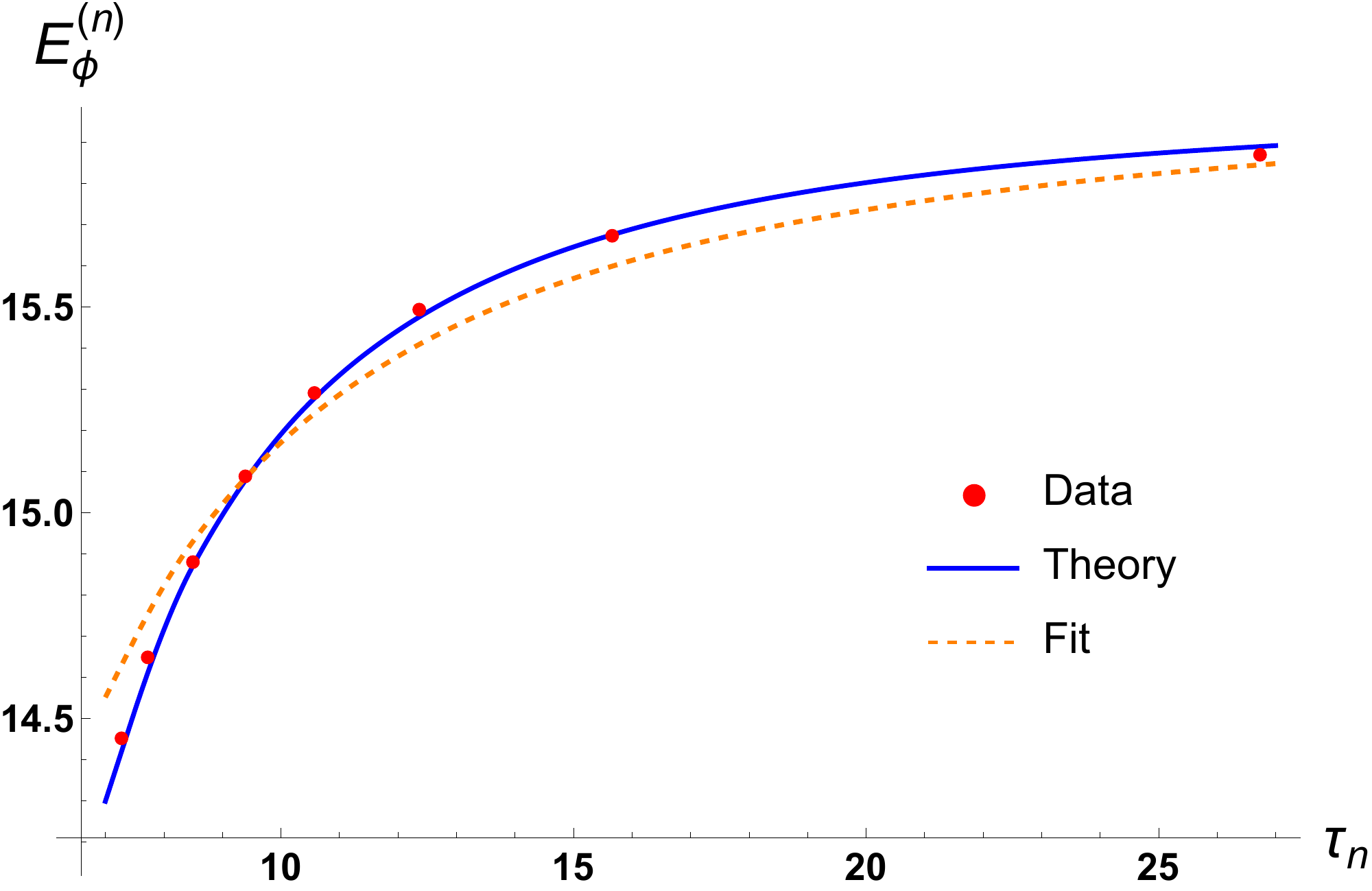}
    \caption{
    Plot of the plateau energy for $\phi$ ($E_\phi^{(n)}$) as a function of half period $\tau_n$ from simulation (red), from theory~\eqref{eq:ephihalfprd} (solid blue) and the analytical fit~\eqref{eq:Ephivstau_fit} (dashed orange). The parameters for the simulation points correspond to the solid blue curve in Fig.~\ref{fig:vcomp_bkrxn}. They are $L=100$, $N=500$, $v=0.1$, $\mu = 0.1$, $\lambda=0.3$, $m_{\rm phys}=1$, $\kappa=1$ and $t_0=-100$.
    }
    \label{fig:breather_vp1}
\end{center}
\end{figure}

In cases where the kink and antikink scatter to form a bound state, we observed that there is a cascade of bursts of radiation at decreasing time intervals. In this subsection we focus on the blue curve in Fig.~\ref{fig:vcomp_bkrxn} corresponding to the choice of parameters $\lambda=0.3$, $\mu=0.1$ and $v=0.1$. This case features particularly well-defined energy plateaus and lends itself to further analysis.  We plot $E_{\phi}^{(n)}$, the energy after the $n^{\rm th}$ burst as a function of $n$ in Fig.~\ref{subfig:delEn}
and find the fit
\be
E_{\phi}^{(n)} \simeq E_{\phi}^{(0)}-0.20\ n\,,
\label{DeltaEn}
\ee
where $E_{\phi}^{(0)}\simeq 16.04$ is the initial energy.
Using Fig.~\ref{fig:plateau_n_vp1} we can also characterize the time intervals $\tau_n$ between the $n^{\rm th}$ and $(n+1)^{\rm th}$ bursts (or between two successive zeros of $\phi(t,x=0)$), as shown in Fig.~\ref{subfig:taun}, to get the fit
\be
\tau_n \simeq \tau_1 n^{-0.66}\,,
\label{taun}
\ee
where $\tau_1= 26.75$ is the first burst interval.
We can then eliminate $n$ to obtain the relation between $E_\phi^{(n)}$ and $\tau_n$,
\be
\label{eq:Ephivstau_fit}
E_{\phi}^{(n)} \simeq E_{\phi}^{(0)} - 0.20 \left ( \frac{\tau_1}{\tau_n} \right )^{1.52}\,.
\ee
Using the exact solution for the sine-Gordon breathers, the relation between the energy $E_\phi$ and the half-period $\tau$ is 
\beq
\label{eq:ephihalfprd}
E_\phi=16 \sqrt{1-{\frac{\pi^2}{\tau^2}}}\,.
\eeq
Fig.~\ref{fig:breather_vp1} shows an excellent agreement of the numerical data with the analytic sine-Gordon prediction for the relation between the plateau energy and the half-period, showing that the $\phi$ field decay proceeds discretely through a series of breather states.
This decay of the energy continues for a finite number of bursts and then, at a critical time, the system rapidly decays into a new, more stable, oscillon-like phase (which we study in more detail in the next subsection).

\subsection{Formation of a long-lived oscillon}
\label{subsec:cslp9}

\begin{figure}
    \begin{center}
         \subfloat[] {\includegraphics[width=0.45\textwidth]{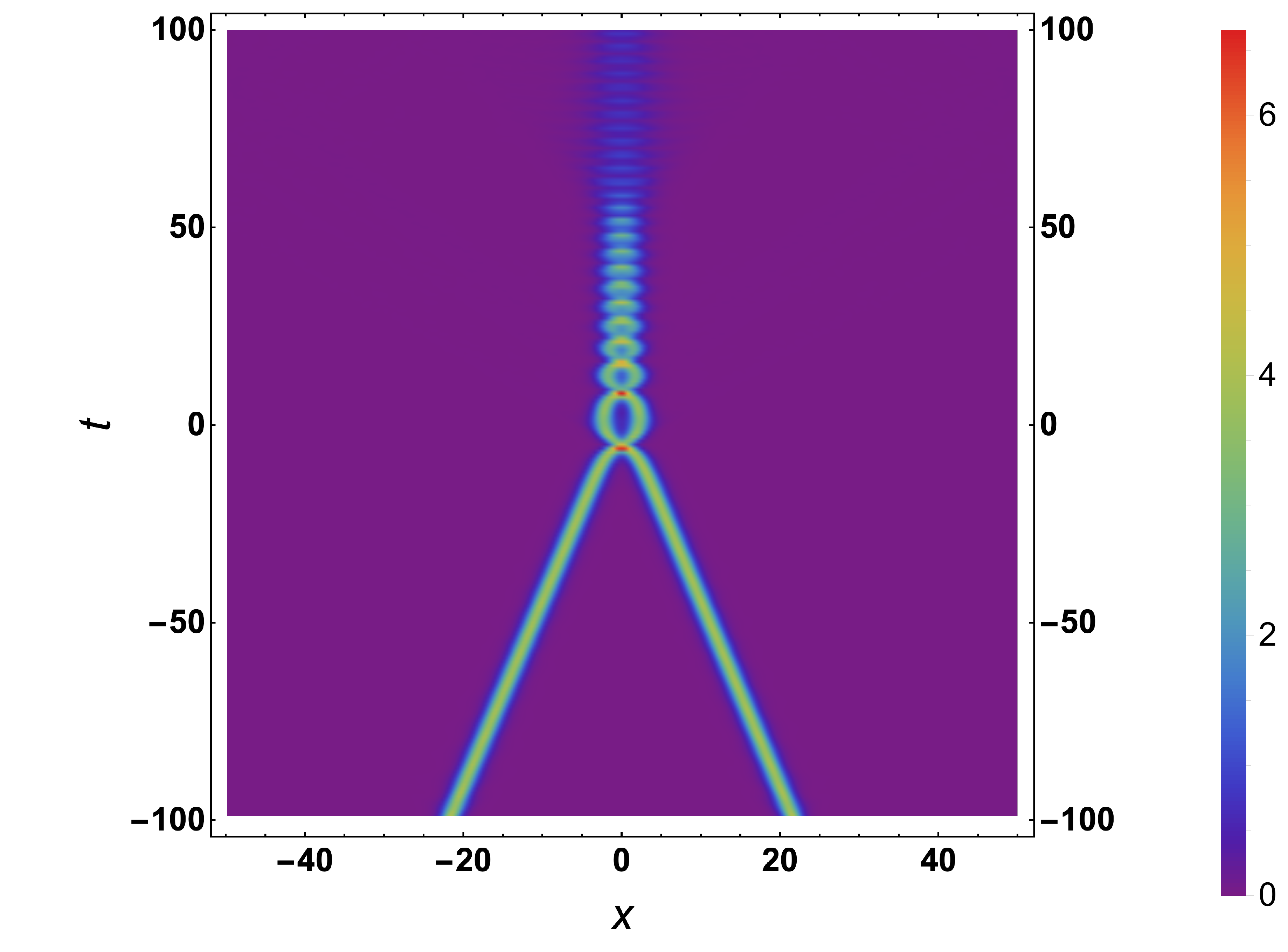}\label{subfig:phiendenlp9}}\hfill%
        \subfloat[] {\includegraphics[width=0.46\textwidth]{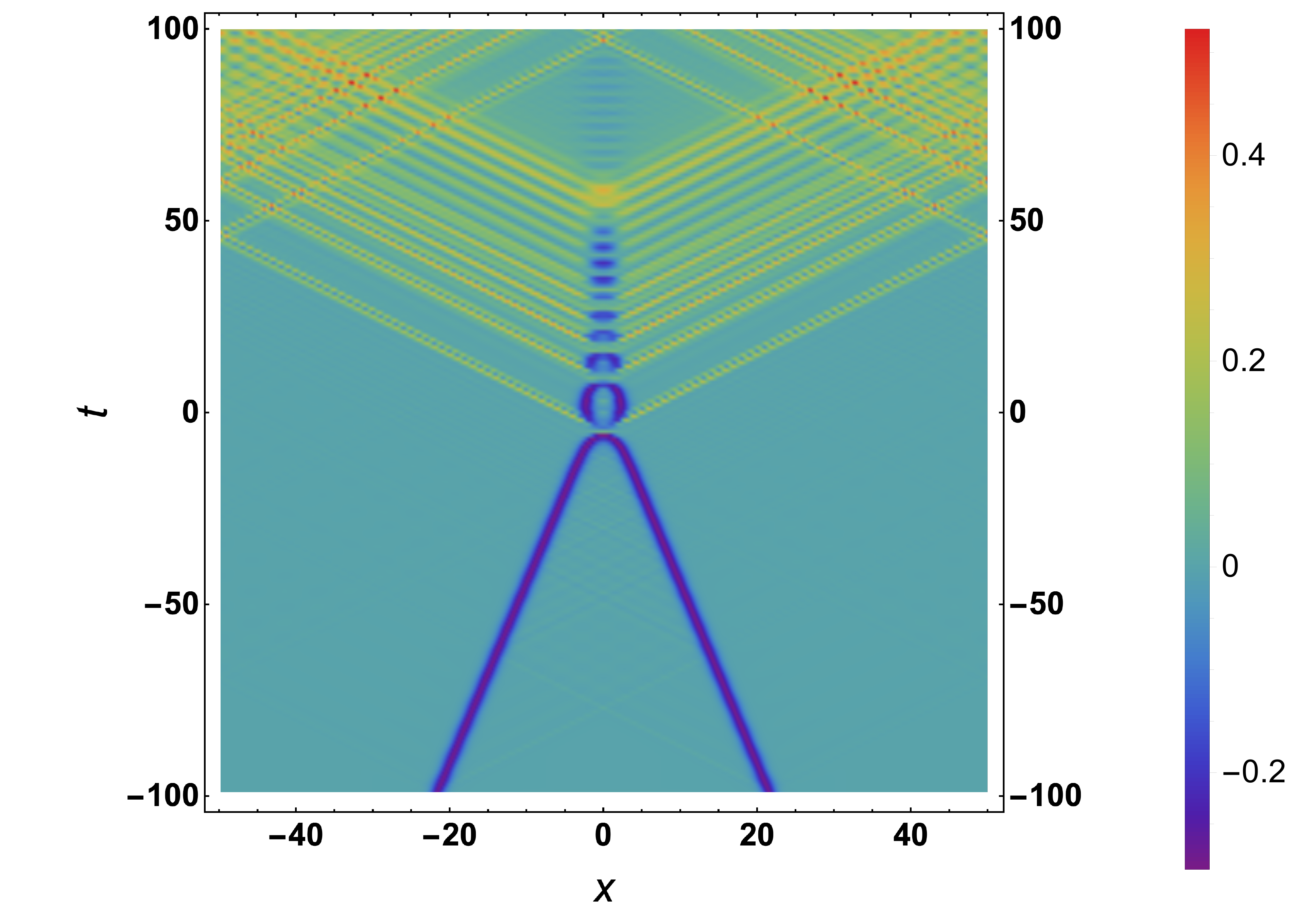}\label{subfig:psiendenlp9}}
        \\
        \subfloat[] {\includegraphics[width=0.45\textwidth]{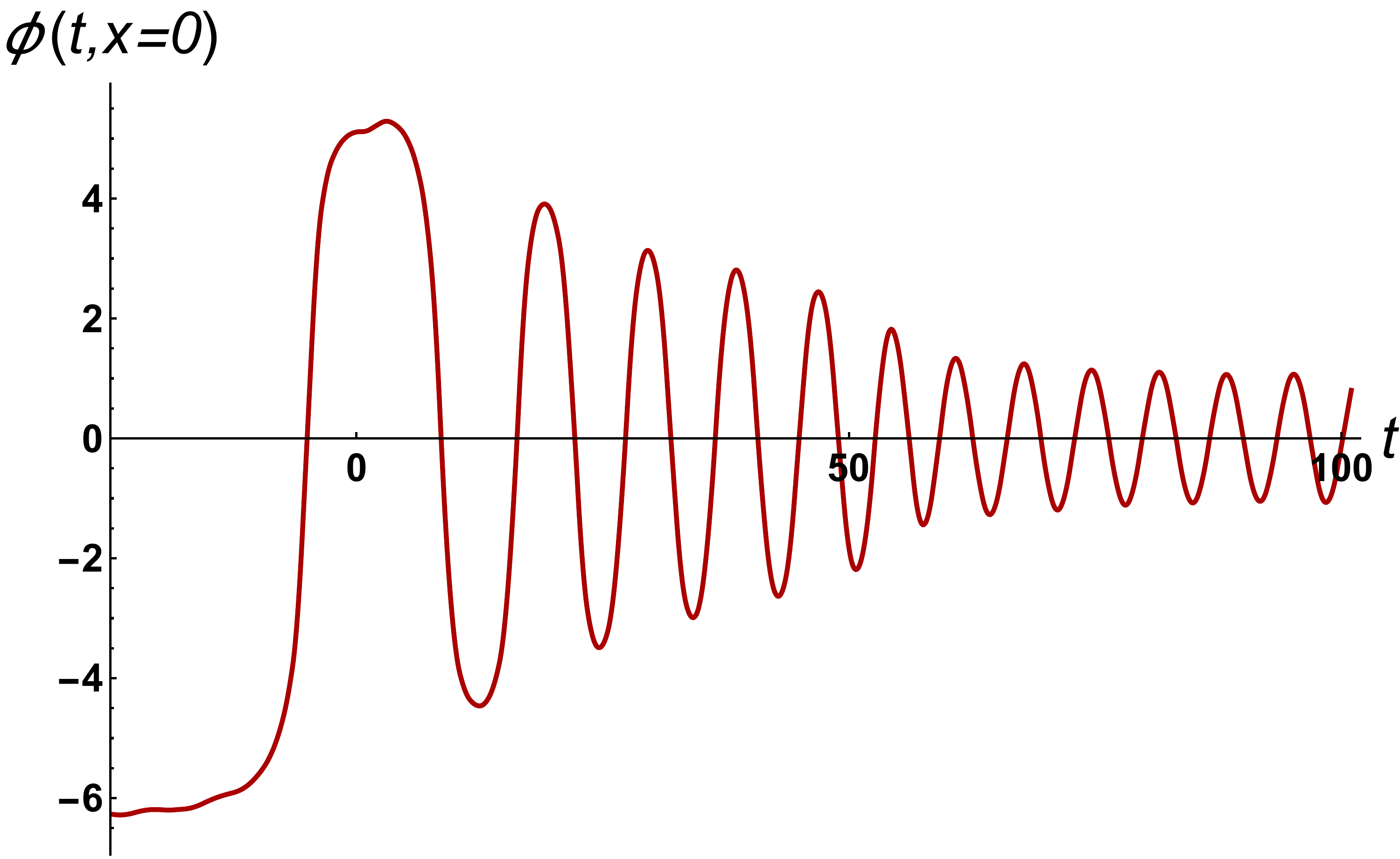}\label{subfig:phi0lp9}}\hfill%
        \subfloat[] {\includegraphics[width=0.45\textwidth]{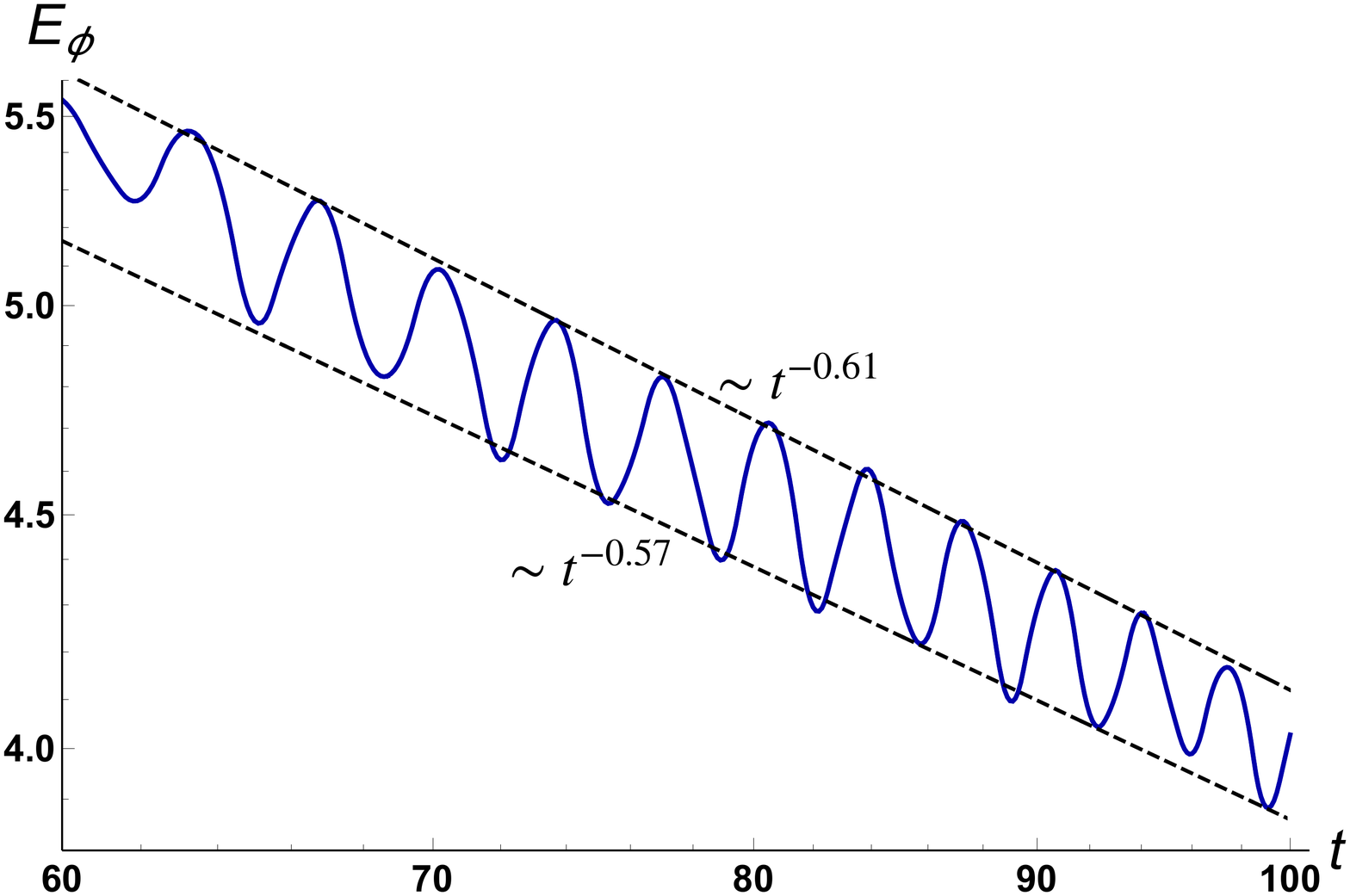}\label{subfig:enphilp9}}
        \caption{Different observables for $\lambda=0.9$ to highlight the formation of a long-lived oscillon: (a) Time evolution of the energy density in $\phi$; 
        (b) Time evolution of the energy density in $\psi$ ($\rho^{(R)}_\psi$); (c) Time evolution of $\phi (t,x=0)$;
        (d) Log-log plot of the late-time evolution of the energy in the $\phi$ ($E_\phi$).
        The parameters
        are $L=100$, $N=500$, $v=0.2$, $\mu = 0.1$ and $m_{\rm phys}=1$, $\kappa=1$ and $t_0=-100$. The collision happens at $t=0$. The animation corresponding to this case can be found at \href{https://sites.google.com/asu.edu/mainakm/animations\#h.oaaca9ogoc42}{https://sites.google.com/asu.edu/mainakm}. 
        }
        \label{fig:casestd_lp9}
    \end{center}
\end{figure}

\begin{figure}
\begin{center}
\includegraphics[width=0.65\textwidth,angle=0]{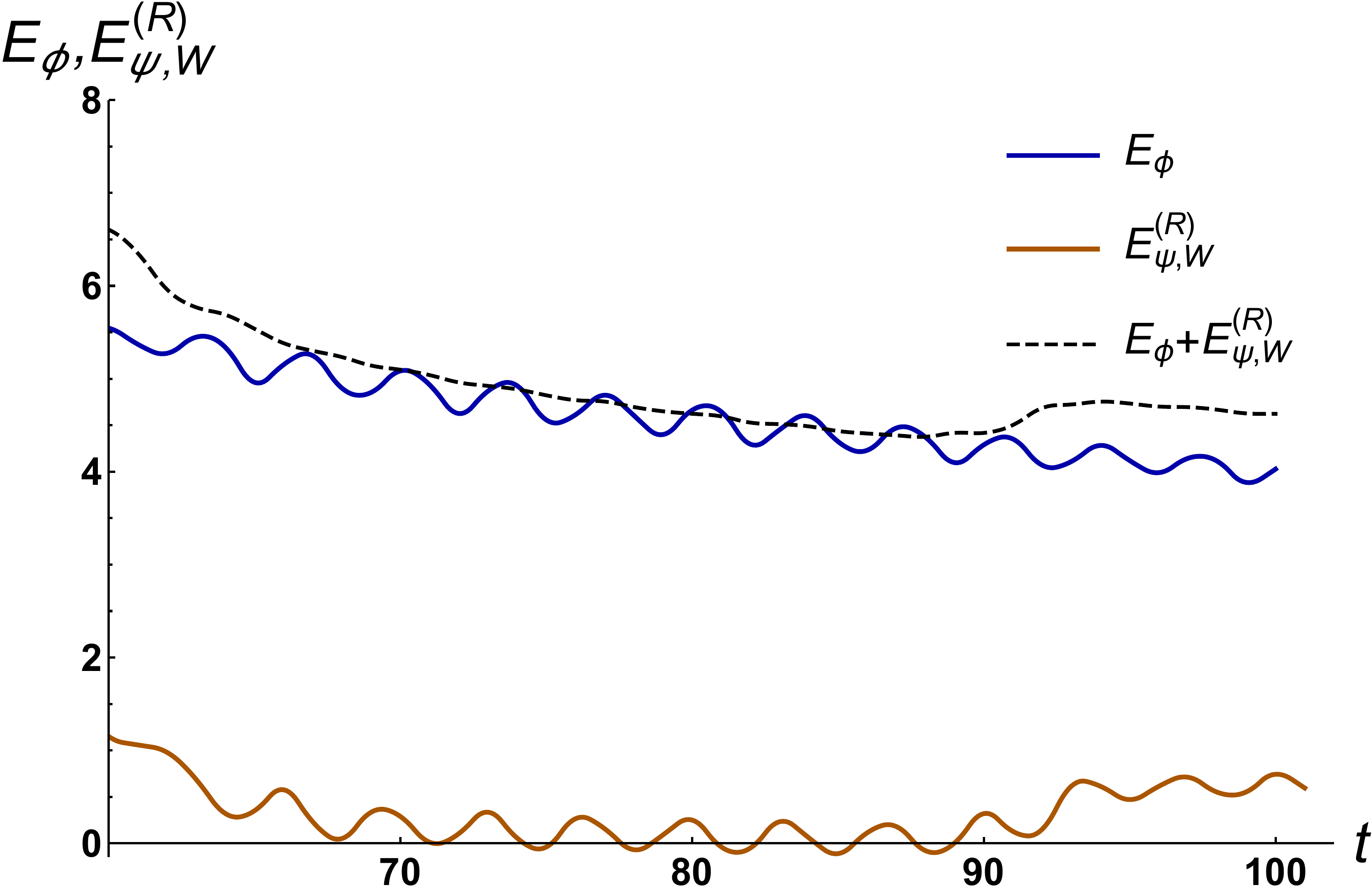}
\caption{\label{fig:phi_psi_encomp} The energy in $\phi$ ($E_\phi$) in solid dark blue, and the renormalized energy in $\psi$ within a window ranging from $x= -5$ to $x=5$ ($E_{\psi,W}^{(R)}$) in solid brown,  as a function of time for $\lambda=0.9$. The sum of the two is denoted by the dashed black line. The parameters
are $L=100$, $N=500$, $v=0.2$, $\mu = 0.1$, $m_{\rm phys}=1$ and $t_0=-100$. Notice that the slight increase in $E_{\psi,W}^{(R)}$ for $t \gtrsim 90$ is due to the emitted radiation reentering the window because of the periodic
boundary conditions.
}
\end{center}
\end{figure}

In this subsection we focus on the particular case $v=0.2$, $\mu=0.1$ and $\lambda=0.9$ (depicted in the blue curve of Fig.~\ref{fig:lcomp_bkrxn}) which prominently features a long-lived oscillon phase and allows us to study its formation and decay. In Figs.~\ref{subfig:phiendenlp9} and~\ref{subfig:psiendenlp9} we show the time evolution of the energy densities in $\phi$ and $\psi$ respectively. In both cases, we see a qualitative change in the appearance of the breather-like structure occurring around $t=70$: the spatial extension of the object ceases to decrease and its energy density undergoes small oscillations in amplitude (visible on both plots).
Fig.~\ref{fig:phi_psi_encomp} shows that the oscillations in the energy of $\phi$ are accompanied with out-of-phase oscillations in the localized energy of $\psi$ around the origin.
As can be seen in Fig.~\ref{subfig:psiendenlp9} the onset of this new regime is preceded by a large burst of radiation that almost completely turns off
at $t \sim 70$. 
Fig.~\ref{subfig:phi0lp9} shows the amplitude of the value of the $\phi$ field at the center of the lattice as a function of time. This is in fact  a good measure of the $\phi$ field profile amplitude. The same qualitative change of behavior is observed on this plot: starting with the moment of collision, the $\phi$ field profile undergoes oscillations of decaying amplitude until a new quasi-stable oscillatory regime is reached around $t=70$. Although, the amplitude of oscillation continues to decrease, it does so at a very slow rate which justifies the long-lived oscillon nomenclature.
Finally, Fig.~\ref{subfig:enphilp9} is simply a zoomed in version of the blue curve in Fig.~\ref{fig:lcomp_bkrxn} allowing us to fit a power law to the envelope of $E_\phi$ within the long-lived oscillon regime. We find a $\sim t^{-0.6}$ power law decay. This is to be contrasted to the approximately linear decay occurring during the  immediately preceding phase, where, on average, the energy decreases approximately linearly in time with a slope of $-0.16$.
Given this power-law decay, this object may not be a true ``oscillon'', since oscillons have been seen to remain highly stable for hundreds or thousands of oscillation times. In order to distinguish it from the more strongly radiating breather state that precedes it, we will adhere to the ``oscillon'' nomenclature for the remainder of this work.

\begin{figure}
    \begin{center}
         \subfloat[] {\includegraphics[width=0.45\textwidth]{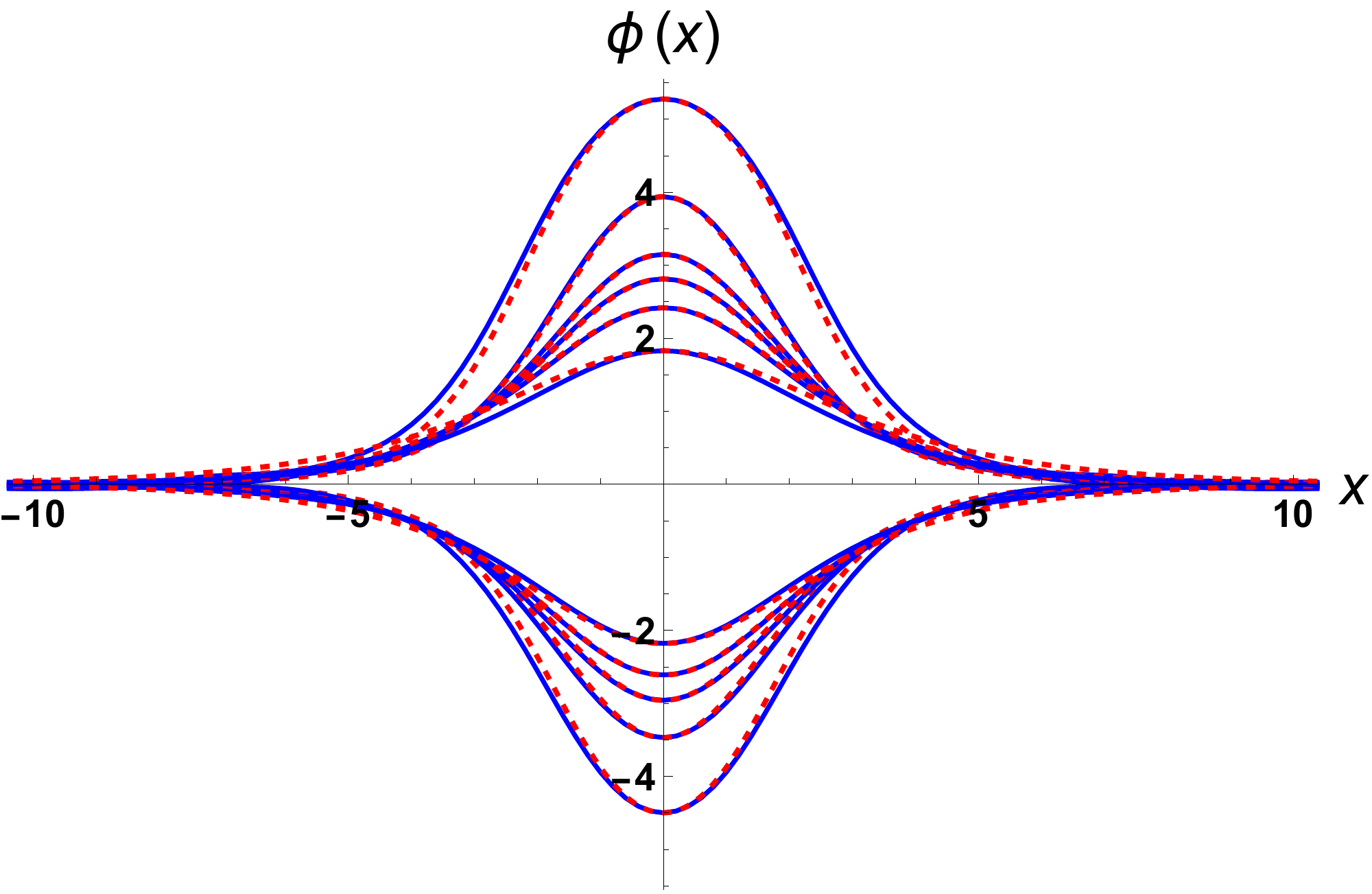}\label{subfig:breathers}}\hfill%
        \subfloat[] {\includegraphics[width=0.46\textwidth]{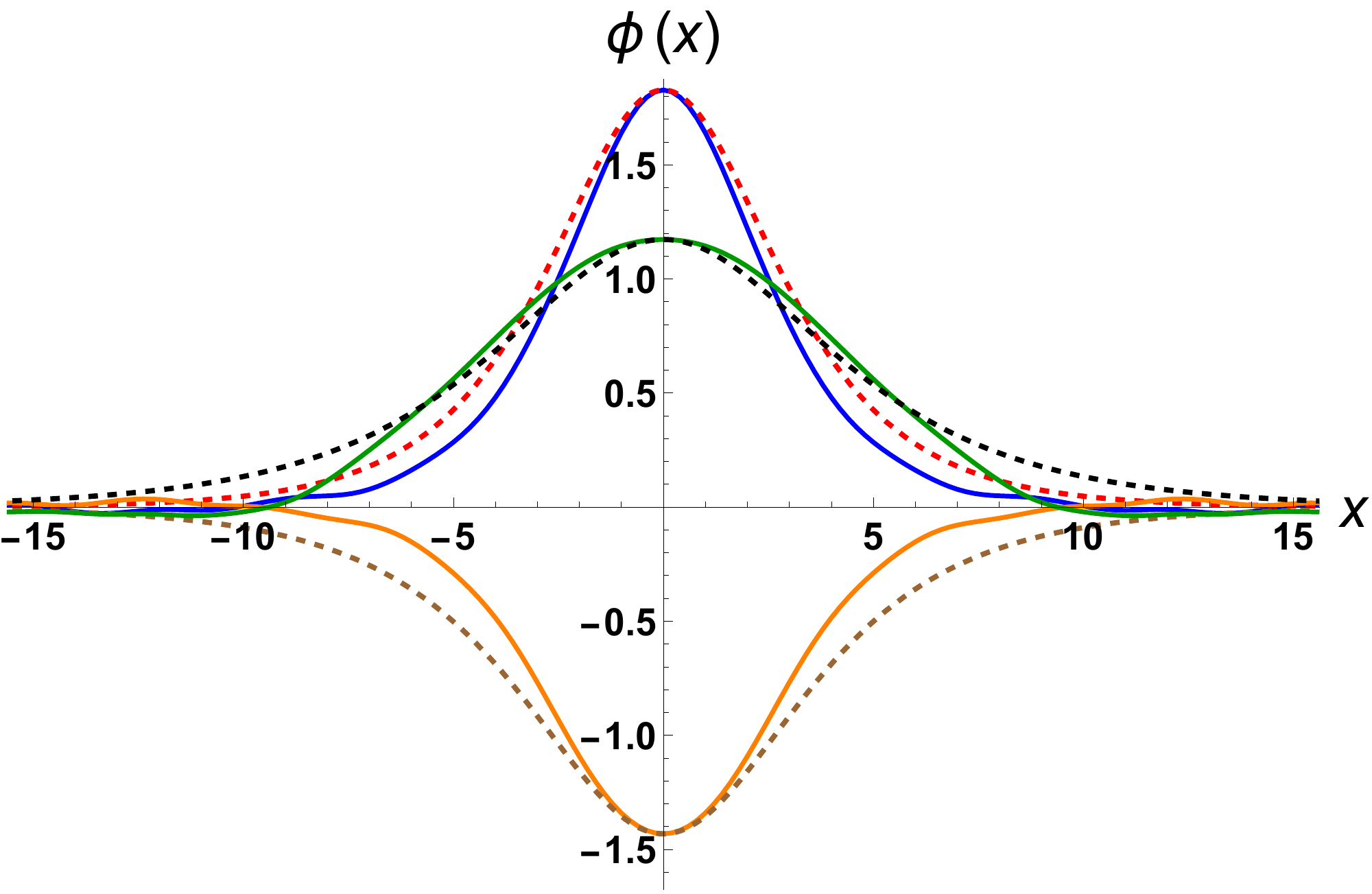}\label{subfig:oscillons}}
        \\
         \subfloat[] {\includegraphics[width=0.45\textwidth]{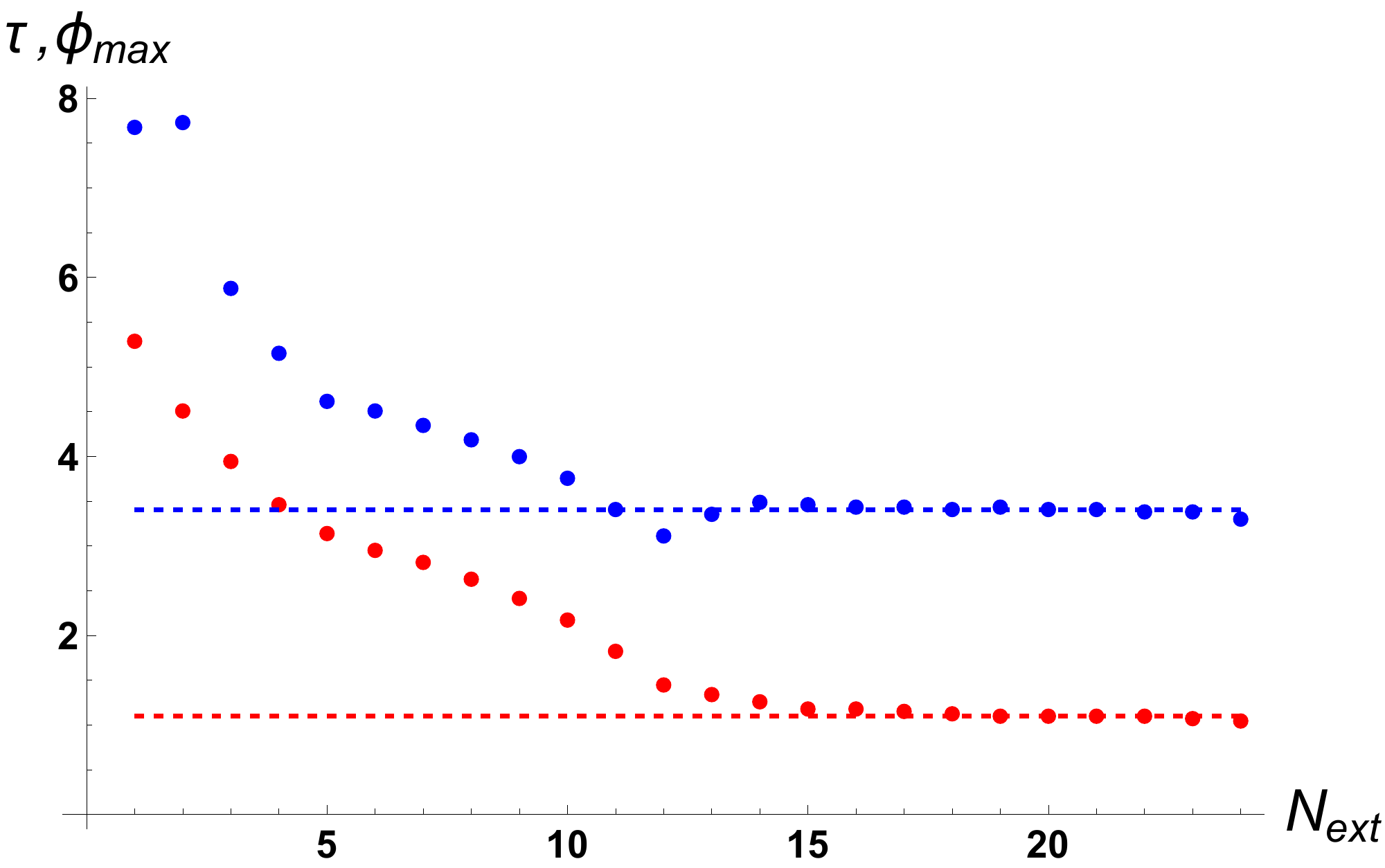}\label{subfig:extrema}}\hfill%
        \subfloat[] {\includegraphics[width=0.46\textwidth]{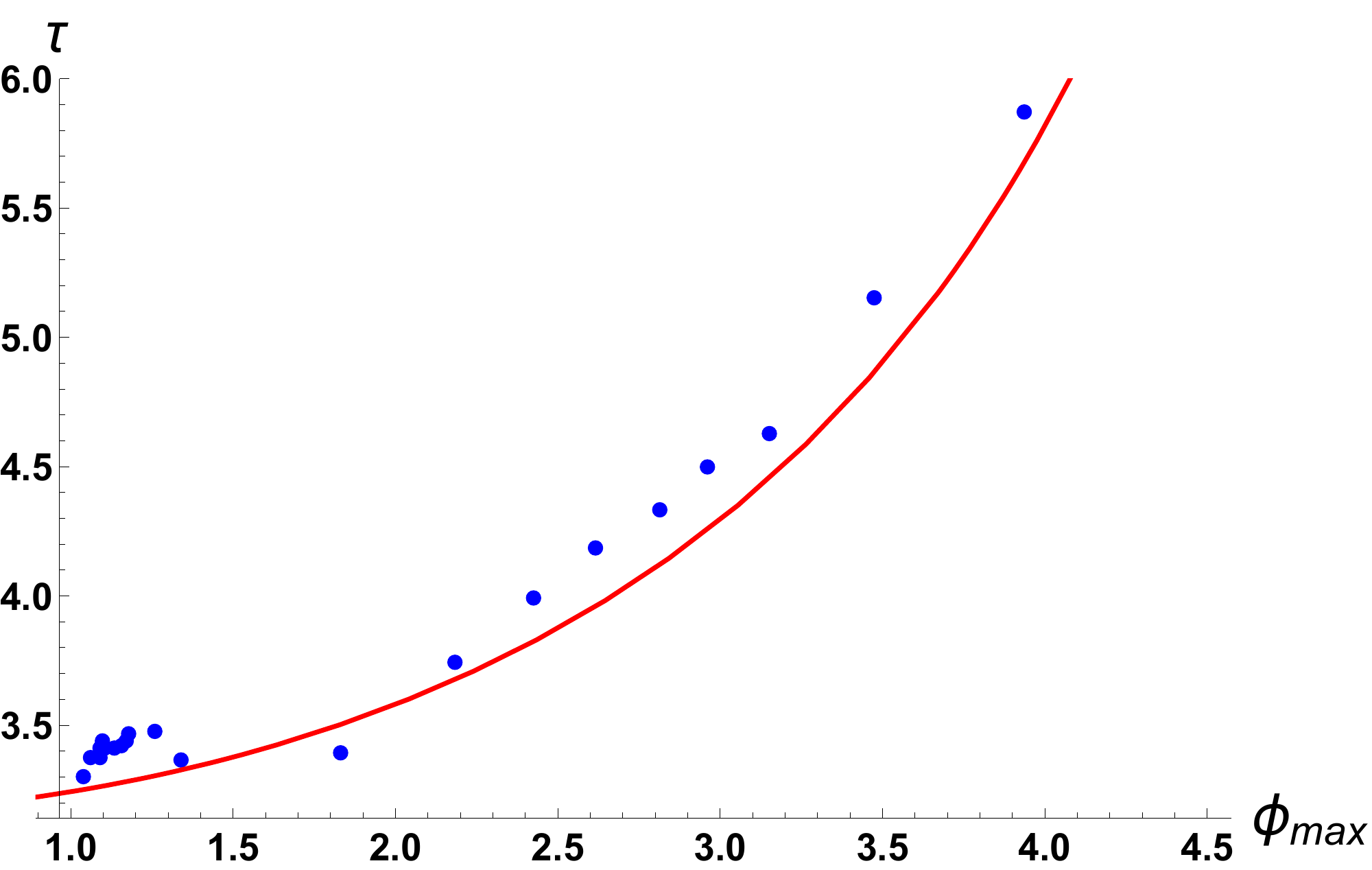}\label{subfig:periodamplitude}}
        \caption{
        (a) The numerically derived profiles of $\phi$ at the first 11 extrema  of Fig.~\ref{subfig:phi0lp9} (blue). The red-dashed curves correspond to the breather profile of Eq.~\eqref{eq:breathershape} with the frequency parameter $\omega$ chosen such that we recover the maximum amplitude $\phi(x=0)$. We see excellent agreement for almost all curves;
        (b) The shape of $\phi$ at the 11$^{\rm th}$, 12$^{\rm th}$ and 15$^{\rm th}$ extremum (blue,orange and green respectively). The red, brown and black curves correspond to the breather shape of  Eq.~\eqref{eq:breathershape};
        (c) The half-period $\tau$ (blue) and maximum central field amplitude $\phi_{\rm max}$ (red)  as a function of the extremum number $N_{\rm ext}$, as in Fig.~\ref{subfig:phi0lp9};
        (d) The half-period as a function of amplitude extracted from the simulation (blue dots) and computed using the analytical breather solution (red curve).
        }
        \label{fig:breathervoscillon}
    \end{center}
\end{figure}

Fig.~\ref{fig:breathervoscillon}  further showcases the behavior of the field $\phi$ in the two regimes. We see that  for early times, when there is significant energy loss every time $\phi$ goes through zero, the maxima of the field $\phi$ match almost exactly to the breather profile of the fully classical Sine-Gordon equation. However, during the second, slowly radiating, part of the evolution, we can see a non-negligible difference between the numerical profile for $\phi$ and that of a sine-Gordon breather. We can thus conclude that, while the early time evolution can be thought  of a series of breathers, with the system ``jumping'' from one to another every time $\phi$ crosses zero, the late time evolution exhibits a deformed breather-like structure (which we denote as an oscillon), which is much less radiating.
The time between successive $\phi$ zero-crossings matches rather well with the half-period of Sine-Gordon breathers, as shown in Fig.~\ref{subfig:periodamplitude}. We also see that both the frequency and the amplitude of the oscillon-like structure remain almost constant.

One potential source of worry is whether the semi-classical approximation remains valid long enough for the oscillon to form. In other words, can the $\phi$ field still be treated as a classical field, even after a significant amount of its energy has dissipated via quantum radiation? As mentioned in Sec.~\ref{sec:setup}, the approximation is valid as long as $E_\phi$ is larger than $E_\psi^{(R)}$ minus the energy radiated away, or in other words, as long as the classical energy in the $\phi$ field is larger than the quantum energy in the clouds of $\psi$ particles ``dressing up'' the kink-antikink pair. As shown in Fig.~\ref{fig:phi_psi_encomp} we have checked that this is indeed the case for the whole duration of the simulation and in particular during the onset of the long-lived oscillon regime.

It would be interesting to understand the profound reasons behind the formation of the long-lived oscillon. 
However, within the time constraints of our simulation, which we can only evolve until $t\simeq 100$ (see Section~\ref{sec:conclusions}), only a small fraction of the parameter choices allow us to reach this oscillon-like final state. 
Improved numerics, able to simulate this late-time long-lived state, will allow us to compare it decay rate to analytical estimates \cite{Hertzberg:2010yz}, thereby further elucidating the truly quantum mechanical nature of the slow decay.
Overall, it is important to understand the formation and eventual decay of this slowly radiating object and we will return to this point in the future.

\section{Discussion and conclusion}
\label{sec:conclusions}

In this work we were able to simulate kink-antikink scattering in a model where a classical sine-Gordon field $\phi$ (the background) is coupled to a quantum field $\psi$ (the quantum radiation bath) via an interaction term preserving the discrete shift symmetry. As expected the spacetime dependence of the background kink-antikink configuration triggers particle production in the $\psi$ field which can in turn backreact on the field $\phi$. As detailed in Sec.~\ref{sec:setup} we used the Classical-Quantum Correspondence (CQC)~\cite{Vachaspati:2018llo, Vachaspati:2018hcu} to study this system numerically. When backreaction is taken into account within the semi-classical approximation, the outcome of the scattering is either an outgoing kink-antikink pair with reduced kinetic energy or a breather-like bound state, which continues to radiate. In Sec.~\ref{sec:results} we examine the dynamics of this inelastic process and its dependence of the different parameters of the model. Interestingly enough, in the case where a bound state forms, particle production initially occurs in the form of a cascade of strong radiation bursts, but after some time the system appears to settle in a long-lived, weakly-radiating oscillon configuration. 
Understanding this final state would provide valuable insight into the late-time evolution of realistic systems, like bubble collisions in the early universe. Furthermore, it has been shown that in some cases the presence of spectator fields can enhance the lifetimes of oscillons \cite{Antusch:2015ziz}. Recent work has also described the properties of multi-component oscillons \cite{VanDissel:2020umg}, which could be relevant for understanding this configuration. A  detailed analytical and numerical investigation of these late-time oscillons falls outside the scope of our current work and will be undertaken in the future.

A limitation of our numerical implementation arises from the fact that working on a periodic lattice prevents us from evolving the dynamical system for a time longer than one light-crossing time after the collision. Beyond that point the emitted radiation comes back to interfere with the kink-antikink pair and our results can no longer be fully trusted. To remedy this would either require parallelizing the code so as to increase the size of the lattice while maintaining spatial resolution, or implementing absorbing boundary conditions.
This is necessary for capturing phenomena that develop over long time-scales (like oscillon evaporation) and is currently under development.

One of the other limitations of our numerical setup lies in the choice of the initial conditions given in 
Eq.~\eqref{backreactedZICs} for the $Z_{ij}$ variables. As mentioned in 
Section~\ref{sec:setup}, 
these are technically only valid when the background is in a quiescent state at time $t_0$, {\it i.e.} when its time 
variation can be neglected.
This is approximately true for non-relativistic collisions such as the ones studied here. However, 
when $\gamma$ becomes large, we expect the mismatch of initial conditions to violate adiabaticity 
strongly around $t=t_0$ thus leading to spurious particle production. Fixing this issue would require 
 going beyond the choice of 0-th order adiabatic vacuum, for example by  pasting together 
the known vacuum modefunctions for a boosted kink and for a boosted antikink~\cite{morse1953methods,Vachaspati:2006zz} propagating in the 
opposite direction to obtain more accurate initial conditions for $\bm{Z}$. 
Alternatively one could adiabatically turn on the relative velocity in a well-separated kink-antikink 
pair but this again requires better control over numerical error.
An amusing fact is that collisions of classical solitons have been analyzed in detail in the opposite 
regime, that of ultra-relativistic velocities \cite{Amin:2013eqa}. There, semi-analytic formulas where 
derived,  albeit neglecting any effect of quantum radiation. Extending the CQC into this regime will 
allow us to capture the quantum radiation effects on colliding relativistic domain walls in the early universe.

The methods presented here in the context of the simple model of a sine-Gordon potential in one spatial dimensions
have a wide applicability to a variety of other scenarios.
For example, the analysis carries over directly to kink-antikink collisions in the $\lambda\phi^4$ model \cite{Sugiyama:1979mi, Campbell:1983xu, Anninos:1991un, Dorey:2017dsn}. Moreover its three-dimensional extension would in fact describe domain wall collisions in the early universe in the presence of quantum radiation. This presents an intriguing opportunity to study possible signatures of the interplay between classical and quantum degrees of freedom in a cosmological context.\footnote{Notice also that the kink-antikink configuration discussed here provides an analogy with the particle production phenomena thought to occur during gravitational collapse, black hole production and evaporation.} Furthermore, the formation of oscillons after bubble collisions and their eventual decay will have to be revisited, in order to encompass the backreaction of quantum radiation \cite{Bond:2015zfa}. We plan to address these open questions in future work.

\appendix
\section{Structure of the radiation bursts}
\label{appsec:spect}

We provide a detailed look at the spatial form of the radiation burst that is generated during the first kink-antikink collision. In order to get a ``clean'' signal, we restrict the present analysis to the case where back-reaction of the $\psi$ field on the $\phi$ background is neglected. This is an increasingly good approximation for smaller values of $\lambda$. 

\begin{figure*}
    \begin{center}
         \subfloat[$\mu=0.1$, $\lambda=0.1$, $E^{RB}_{0.1}=0.0161$, $E^{RB}_{0.3}=0.0136$] {\includegraphics[width=0.45\textwidth]{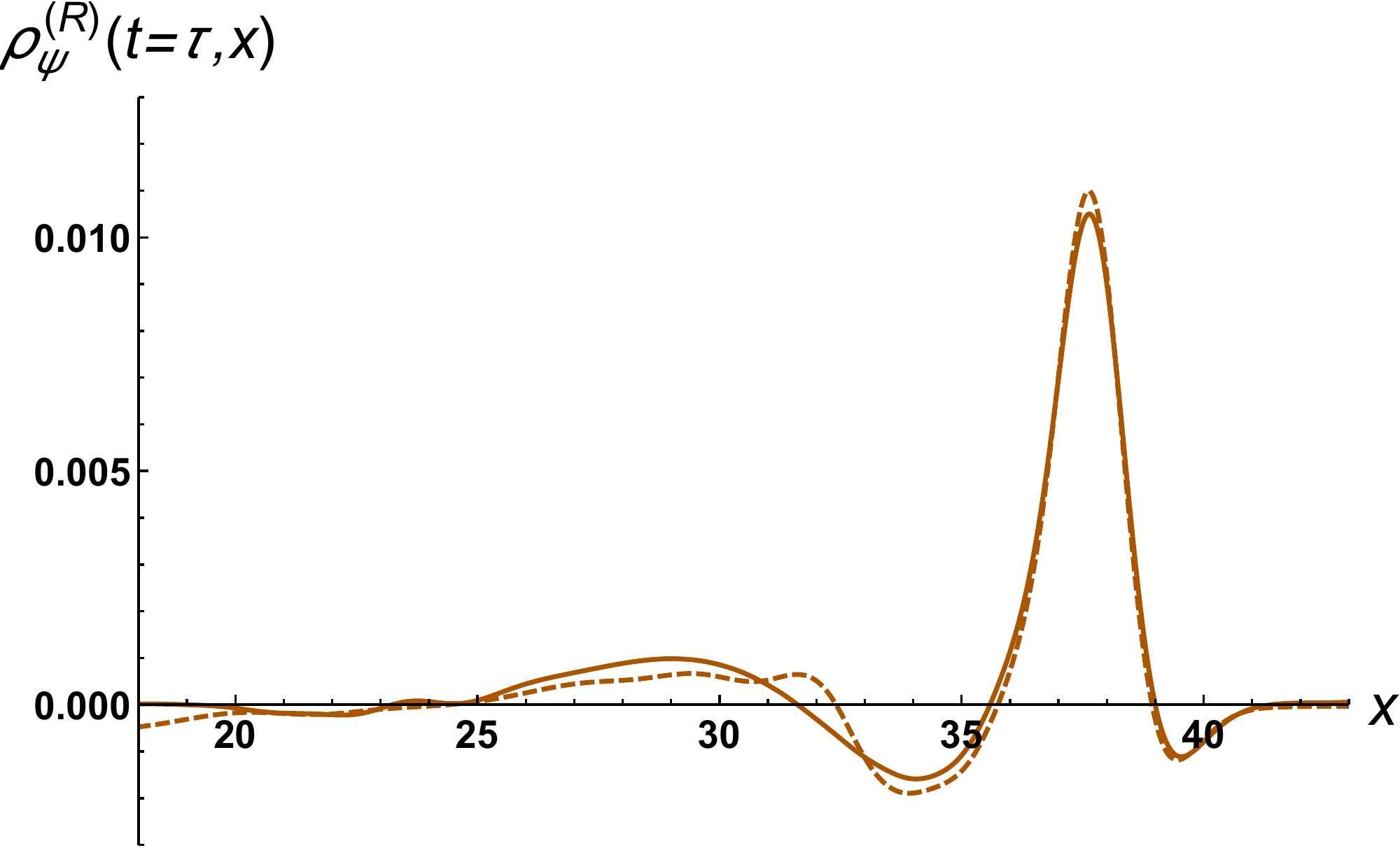}\label{appsubfig:a}}\hfill%
         \subfloat[$\mu=0.1, \lambda=0.9, E^{RB}_{0.1}=0.2799, E^{RB}_{0.3}=0.2698$] {\includegraphics[width=0.45\textwidth]{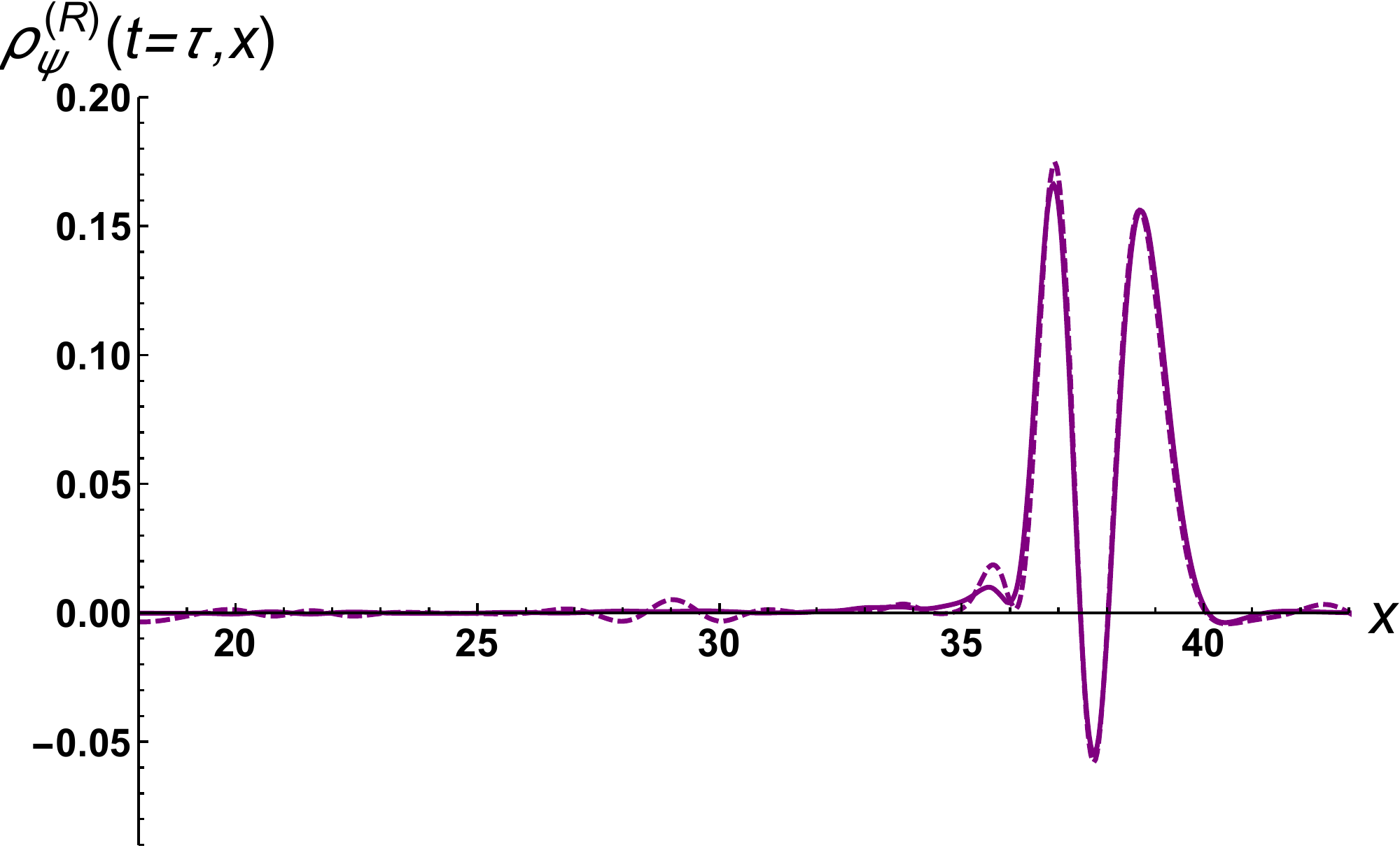}\label{appsubfig:c}}\hfill%
        \\
        \subfloat[$\mu=0.7, \lambda=0.1, E^{RB}_{0.1}=0.0033, E^{RB}_{0.3}=0.0030$] {\includegraphics[width=0.45\textwidth]{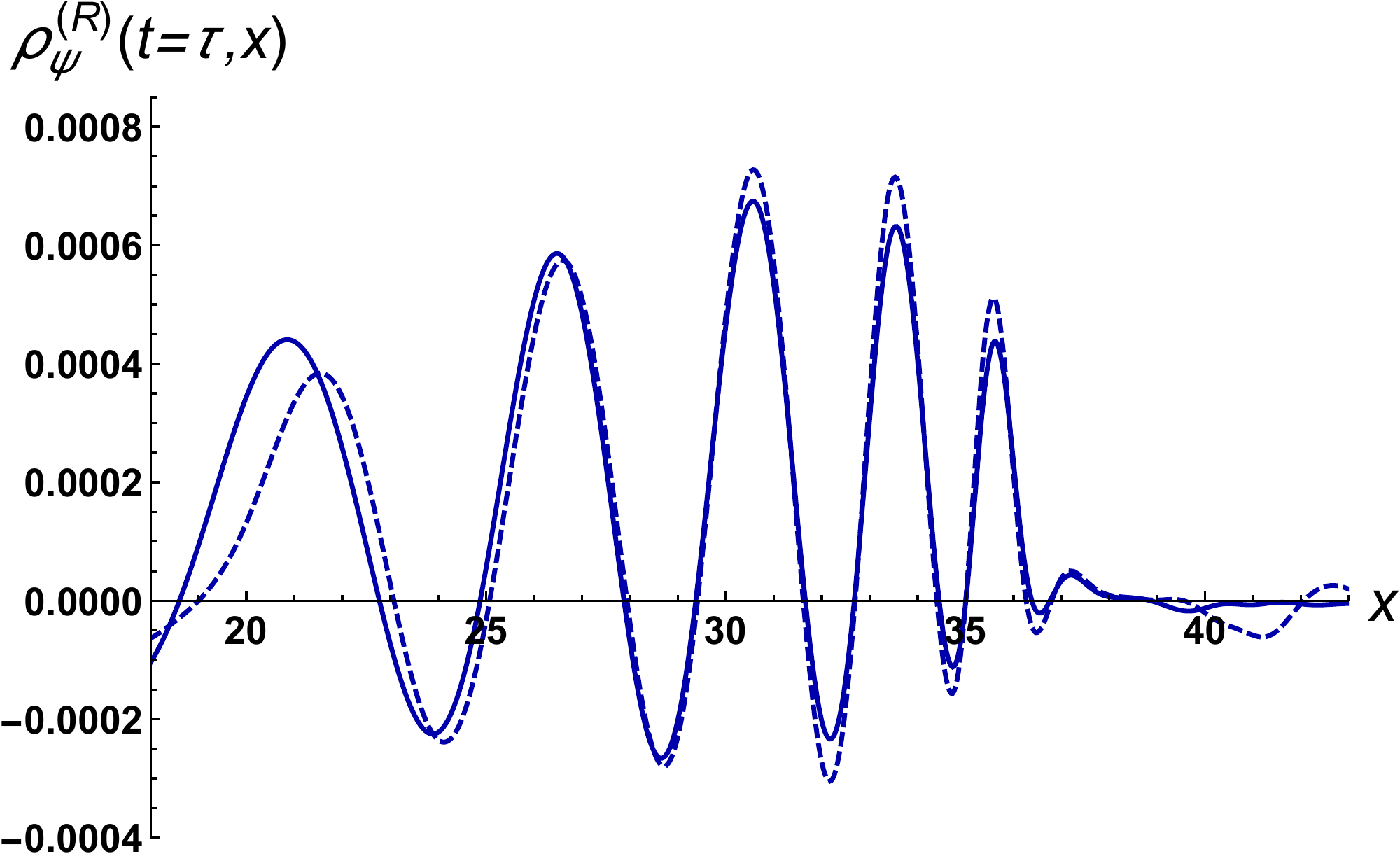}\label{appsubfig:d}}\hfill
        \subfloat[$\mu=0.7, \lambda=0.9, E^{RB}_{0.1}=0.0908, E^{RB}_{0.3}=0.0848$] {\includegraphics[width=0.45\textwidth]{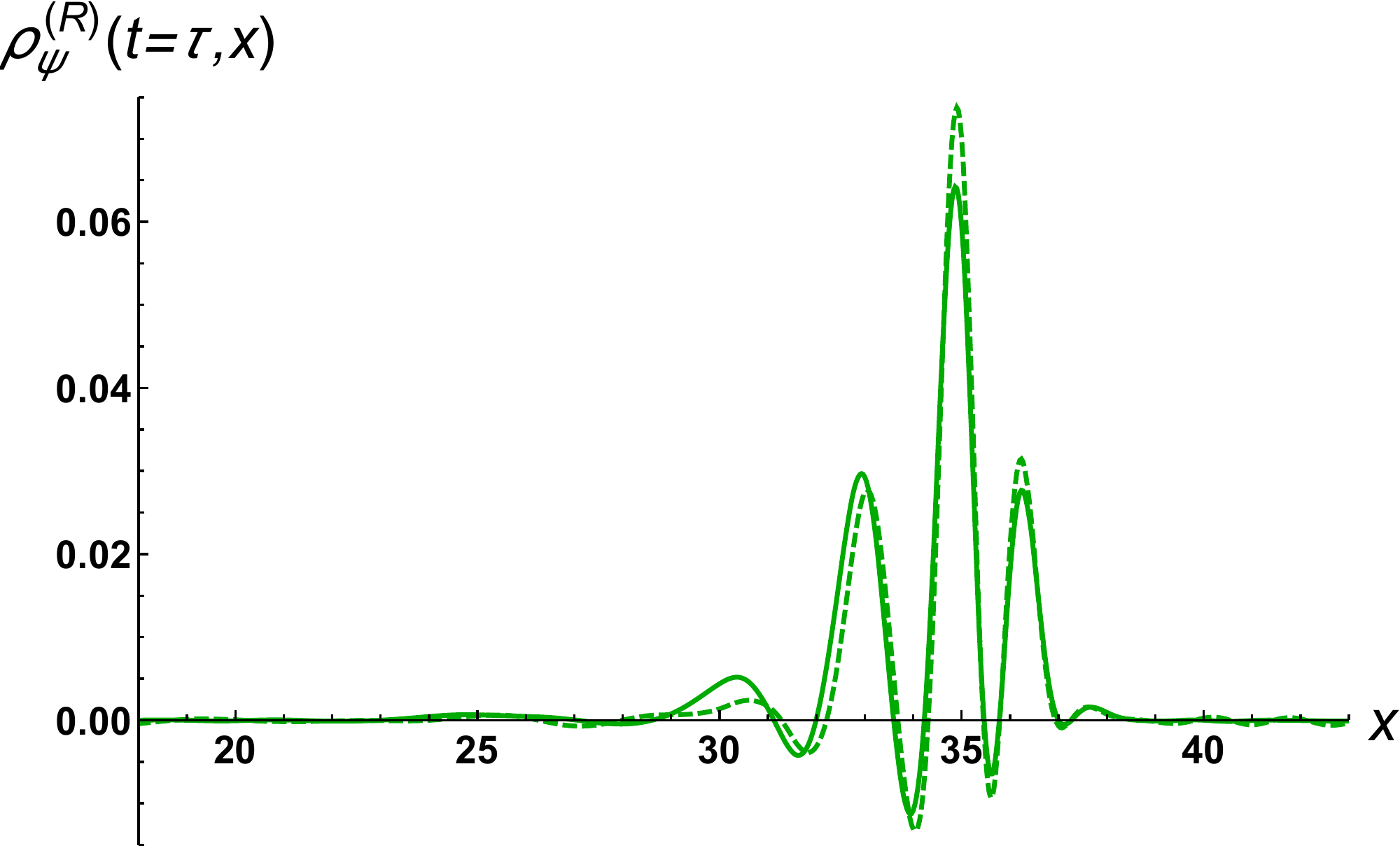}\label{appsubfig:e}}\hfill%
        \caption{Snapshots (taken at time $\tau=40$) of the renormalized energy density of the radiation bursts for different parameters in the case without backreaction. The solid lines are for $v=0.1$ and the dashed lines are for $v=0.3$. $E^{RB}_v$ is the energy in the radiation burst for $v=0.1$ and $v=0.3$. The other parameters
        are $L=100$, $N=500$, $\kappa\rightarrow 0$, $m_{\rm phys}=m=1$ and $t_0=-100$.
        }
        \label{appfig:spectra_plots}
    \end{center}
\end{figure*}

Fig.~\ref{appfig:spectra_plots} shows the resulting radiation bursts, where they are sufficiently far away from the collision region. We choose to present two significantly different values for each of the three parameters of the problem: the velocity ($v=0.1,0.3$), the $\psi$ field mass ($\mu=0.1,0.7$) and the coupling strength ($\lambda=0.1, 0.9$).

An immediate realization concerns the effect of changing the initial kink-antikink velocity $v$. We see that the radiation waveform is almost identical for each pair of $(\mu,\lambda)$ when we change the velocity. 
Furthermore, the energy contained in the radiation burst is almost equal for the two curves in each panel of Fig.~\ref{appfig:spectra_plots}.  Hence, our numerical results given in the main text are corroborated by Fig.~\ref{appfig:spectra_plots}, showing that the quantum radiation emitted during the kink-antikink collision essentially loses the memory of the initial velocity.

A further observation is related to the amplitude and shape of the renormalized energy density $\rho_\psi^{(R)}$ in the radiation burst in each of the four panels of Fig.~\ref{appfig:spectra_plots}, corresponding to the four different combinations of $\lambda$ and $\mu$. Each radiation burst is seen to form a wave packet with a spectral content that is highly dependent on the parameters. Most importantly, we see that increasing $\lambda$ increases the overall energy in $\psi$ radiation. This is to be expected, since  $\lambda$ is the coupling strength between the classical background field $\phi$ and the quantum radiation field $\psi$.
On the contrary, increasing $\mu$ leads to a decreased energy in the radiation burst. This can also be qualitatively understood. As shown in Fig.~\ref{fig:enden_ic}, the initial $\psi$ configuration deviates from the trivial vacuum more for smaller values of $\mu$. Simply put, it is easier to excite a lighter quantum field. Finally, we see that the front of the wavepacket is at a  slightly  larger distance from the origin for $\mu=0.1$ than for $\mu=0.7$. Since all panels correspond to snapshots taken at the same time $\tau = 40$, this translates into a larger velocity of the radiation burst in the case of smaller $\psi$ mass ($\mu$), as one could have guessed from simple kinematical arguments.

\section{Quality of numerics}
\label{appsec:numerics}

\begin{figure}
    \begin{center}
         \subfloat[] {\includegraphics[width=0.46\textwidth]{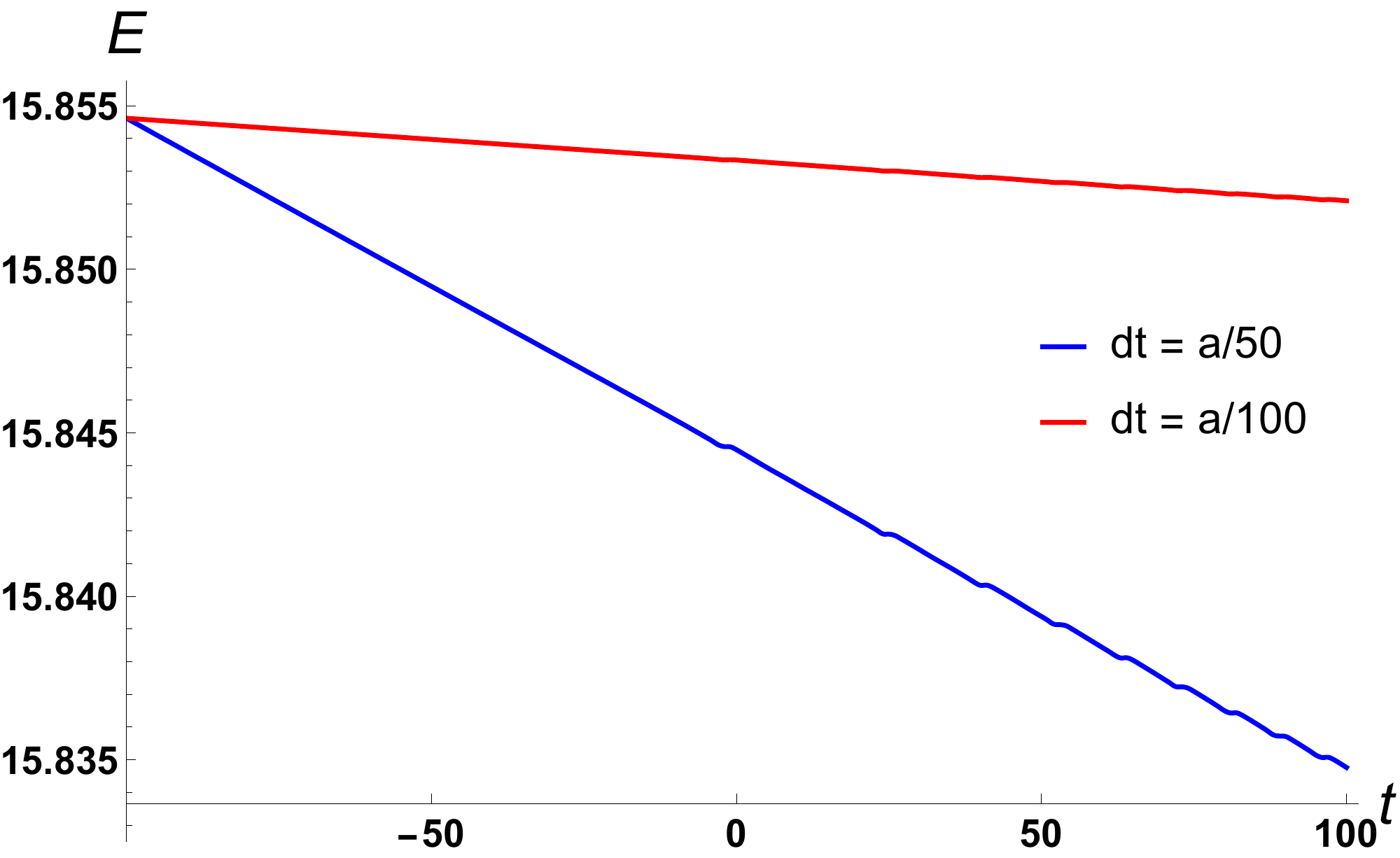}\label{subfig:dtcomp_toten}}\hfill%
        \subfloat[] {\includegraphics[width=0.46\textwidth]{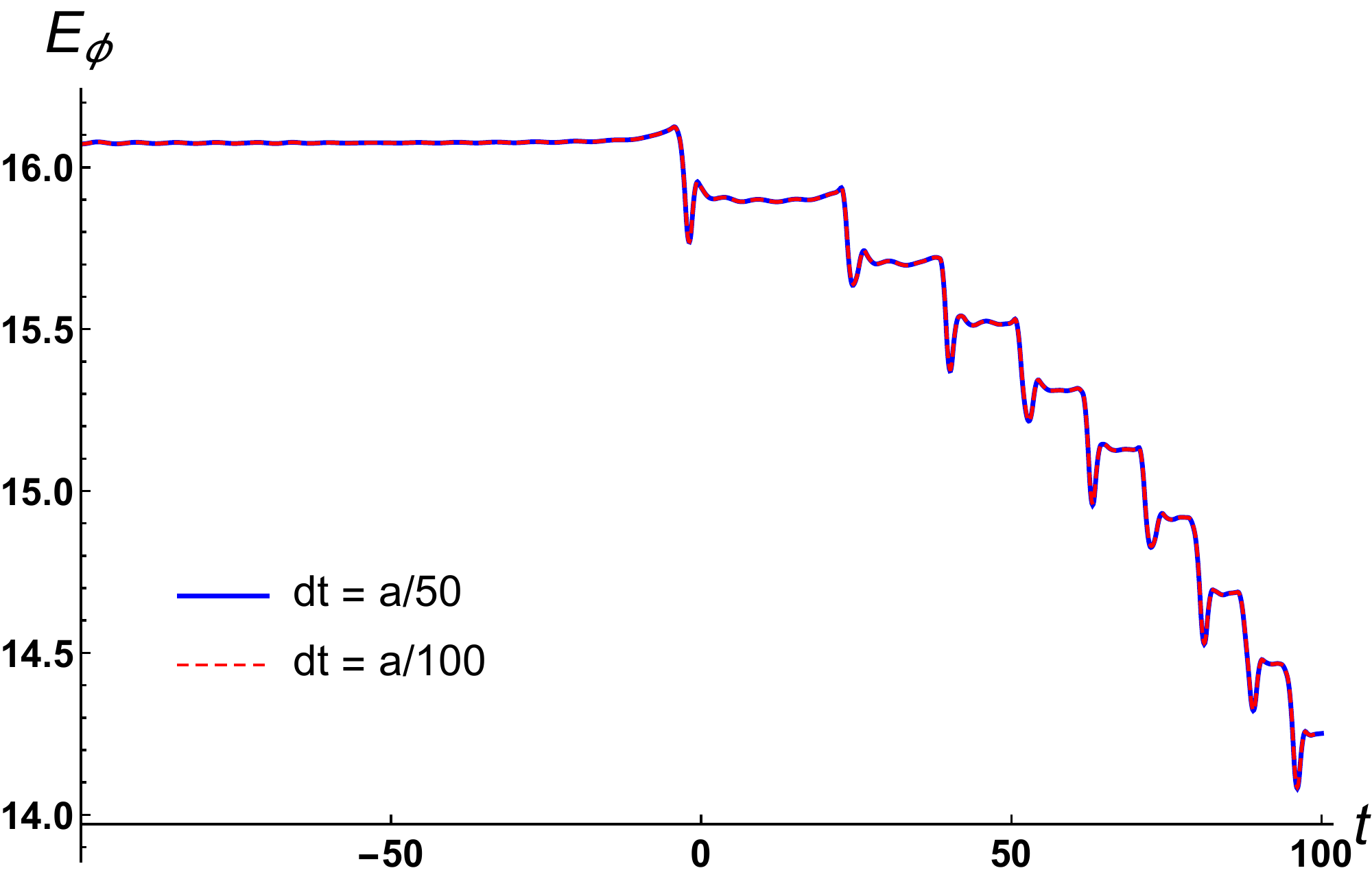}\label{subfig:dtcomp_phien}}
        \caption{(a) Time evolution of the total energy~\eqref{eqn:totalen} for $dt=a/50=0.004$ and $dt=a/100=0.002$.
        (b) Time evolution of the energy in $\phi$ ($E_\phi$) for $dt=a/50=0.004$ (solid blue) and $dt=a/100=0.002$ (dashed red).
        The parameters
        are $L=100$, $N=500$, $v=0.1$, $\mu = 0.1$, $\lambda =0.3$ and $m_{\rm phys}=1$, $\kappa=1$ and $t_0=-100$. The collision happens at $t=0$. Recall that $a=L/N$. These plots illustrate the independence of our results on the choice of time step.
        }
        \label{fig:dtcomp}
    \end{center}
\end{figure}

\begin{figure}
\begin{center}
\includegraphics[width=0.65\textwidth,angle=0]{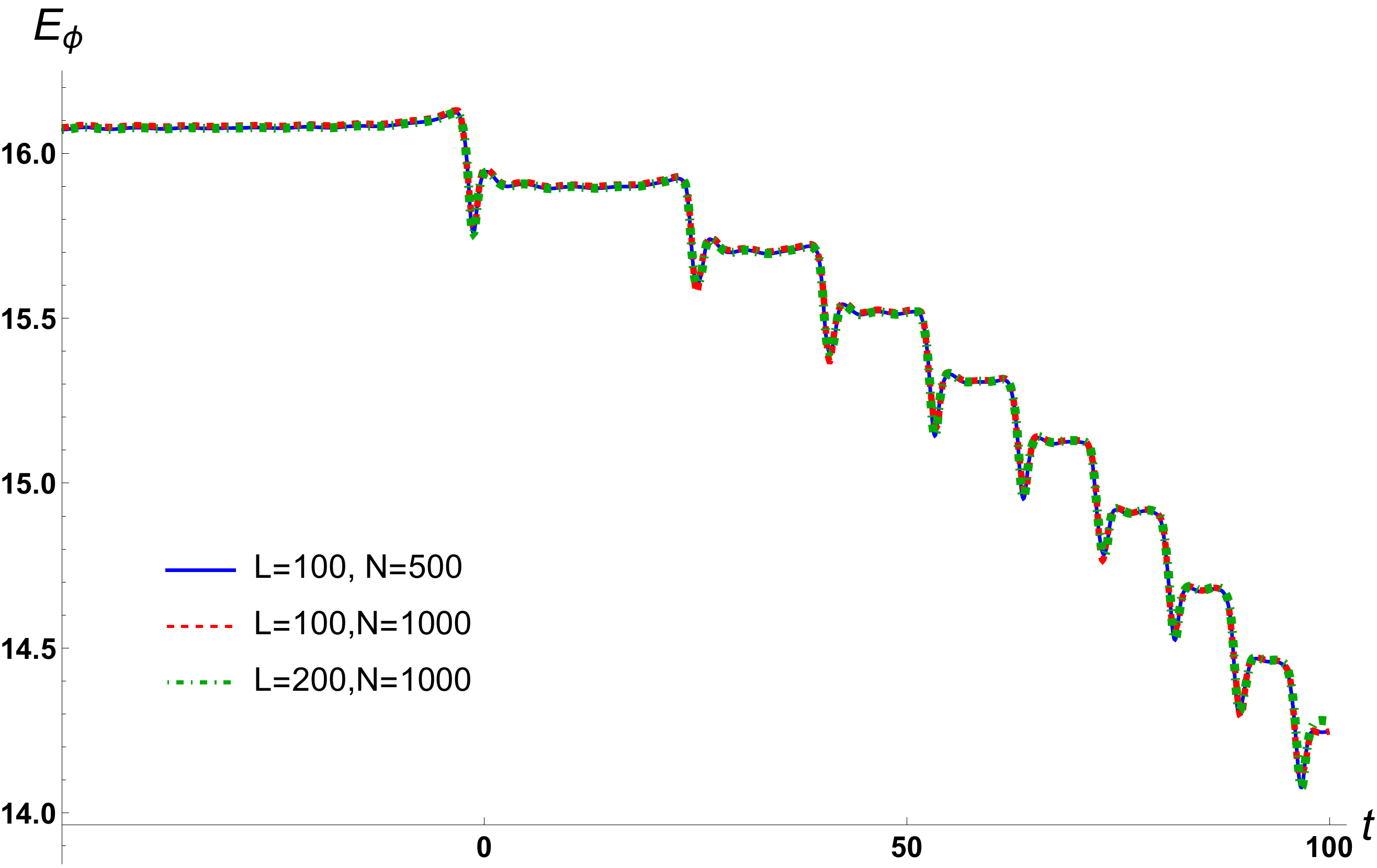}
\caption{\label{appfig:ndep_ldep_check} 
Time evolution of the energy in $\phi$ ($E_\phi$) for different values of $L$ and $N$. The parameters are $v=0.1$, $\mu = 0.1$, $\lambda = 0.3$ and $m_{\rm phys}=1$, $\kappa=1$ and $t_0=-50$. The collision happens at $t=0$. Recall that $a=L/N$ and we have chosen $dt=a/50$. This illustrates the independence of our results on the lattice spacing $a$ and on the size of the box $L$. 
}
\end{center}
\end{figure}

In this appendix, we discuss the quality of the numerics in our simulations. In Fig.~\ref{fig:dtcomp}, we illustrate the independence of the physical observables on the choice of the temporal time step used in the simulation, $dt$. In Fig.~\ref{subfig:dtcomp_toten} we see that the total energy is conserved to an accuracy of $\sim 0.1\%$ over the entire time of evolution (for a time step $dt=0.004$). As expected, energy is conserved to an even better accuracy when dividing the time step by two. However, given the trade-off in computational time we think $dt=0.004$ is sufficient for the task at hand. Our choice is further justified by Fig.~\ref{subfig:dtcomp_phien} where we show the energy in $\phi$ ($E_\phi$) computed for two values of $dt$, $0.004$ and $0.002$. Indeed the two plots superimpose each other. Moreover, Fig.~\ref{appfig:ndep_ldep_check} clearly shows that, at least for the physical observables that we are interested in (i.e. $E_\phi$), decreasing the lattice spacing $a=L/N$ or increasing the size of the box $L$ do not sensibly change our results. In other words, our choices of parameters are good enough and the results shown in the main text are independent of both the UV and the IR cutoffs.

\acknowledgments
We would like to thank Oriol Pujolas for many useful discussions, and Fabio van Dissel for independently verifying the robustness of our numerical results. Computations for this work were performed on the Agave cluster at Arizona State University. The research leading to these results has received funding from the Spanish Ministry of Science and Innovation (PID2020-115845GB-I00/AEI/10.13039/501100011033). IFAE is partially funded by the CERCA program of the Generalitat de Catalunya.
MM was supported by the Fermi National Accelerator Laboratory (Fermilab) Award No. AWD00035045 and the National  Science  Foundation  grant  numbers  PHY-1613708 and PHY-2012195 during this work. 
The work of EIS and GZ  was supported by
a fellowship from ``La Caixa'' Foundation (ID 100010434) and from the European Union's Horizon
2020 research and innovation programme under the Marie Sk\l odowska-Curie grant agreement No
847648. The fellowship code is LCF/BQ/PI20/11760021.
TV is supported by the U.S. Department of Energy, Office of High Energy Physics, under Award DE-SC0019470 at ASU.


\bibliography{sgcqc}
\bibliographystyle{jhep}

\end{document}